\documentclass[lettersize,journal,twoside]{IEEEtran}
\usepackage{amsmath,amsfonts}
\usepackage{amsmath,amssymb,bm,booktabs,multirow,url}
\usepackage{algorithmic}
\usepackage{algorithm}
\usepackage{array}
\usepackage[caption=false,font=footnotesize,labelfont=rm,textfont=rm]{subfig}
\usepackage{textcomp}
\usepackage{stfloats}
\usepackage{url}
\usepackage{verbatim}
\usepackage{graphicx}
\usepackage{cite}
\usepackage{hyperref}
\usepackage{utfsym}
\usepackage{colortbl}
\usepackage{hyperref}
\usepackage{threeparttable}
\usepackage[usestackEOL]{stackengine}
\begin{document}

\title{Learning Switchable Priors for\\Neural Image Compression}
\author{Haotian Zhang, Yuqi Li, Li Li, \IEEEmembership{Member, IEEE,} and Dong Liu, \IEEEmembership{Senior Member, IEEE}
\thanks{Date of current version \today. This work was supported by the Natural Science Foundation of China under Grant 62021001, and also supported by the graphical processing unit (GPU) cluster built by the Multimedia Computing and Communication (MCC) Lab of the Information Science and Technology Institution.

The authors are with the MOE Key Laboratory of Brain-Inspired Intelligent Perception and Cognition, University of Science and Technology of China, Hefei 230093, China (e-mail: zhanghaotian@mail.ustc.edu.cn; lyq010303@mail.ustc.edu.cn; lil1@ustc.edu.cn; dongeliu@ustc.edu.cn). (\textit{Corresponding author: Dong Liu})}}

\markboth{ }
{Zhang \MakeLowercase{\textit{et al.}}: Learning Switchable Priors for Neural Image Compression}

\maketitle

\begin{abstract}
Neural image compression (NIC) usually adopts a predefined family of probabilistic distributions as the prior of the latent variables, and meanwhile relies on entropy models to estimate the parameters for the probabilistic family.
More complex probabilistic distributions may fit the latent variables more accurately, but also incur higher complexity of the entropy models, limiting their practical value.
To address this dilemma, we propose a solution to decouple the entropy model complexity from the prior distributions.
We use a finite set of trainable priors that correspond to samples of the parametric probabilistic distributions.
We train the entropy model to predict the index of the appropriate prior within the set, rather than the specific parameters.
Switching between the trained priors further enables us to embrace a skip mode into the prior set, which simply omits a latent variable during the entropy coding.
To demonstrate the practical value of our solution, we present a lightweight NIC model, namely FastNIC, together with the learning of switchable priors. FastNIC obtains a better trade-off between compression efficiency and computational complexity for neural image compression.
We also implanted the switchable priors into state-of-the-art NIC models and observed improved compression efficiency with a significant reduction of entropy coding complexity.
\end{abstract}
\section{Introduction}
Image compression is a fundamental problem in digital image processing, communications, and computer vision. 
Over the past four decades, significant progress has been made in the development of lossy image compression codecs. Most lossy image compression methods follow the transform coding paradigm \cite{goyal2001transform}, where images are transformed into a latent space for decorrelation, followed by quantization and entropy coding. In traditional image compression codecs, such as JPEG \cite{wallace1991jpeg}, BPG \footnote{BPG is an image coding format based on HEVC intra coding, see \url{http://bellard.org/bpg/}.}, and VVC \cite{bross2021overview}, different modules are manually designed and separately optimized. 

In recent years, neural image compression has demonstrated superior performance by leveraging the capabilities of deep neural networks. Unlike traditional approaches, neural image compression \cite{balle2017end} adopts end-to-end optimization for all components. In a typical scheme of neural image compression \cite{balle2017end}, the encoder applies an analysis transform to the original image, producing latent representations. The latent representations are then quantized and entropy-coded into bitstreams. The decoder recovers the discrete latents and reconstructs the image using a synthesis transform. All modules are jointly optimized during training to minimize the rate-distortion cost.

\begin{figure}[!t]
  \centering \includegraphics[width=0.9\linewidth]{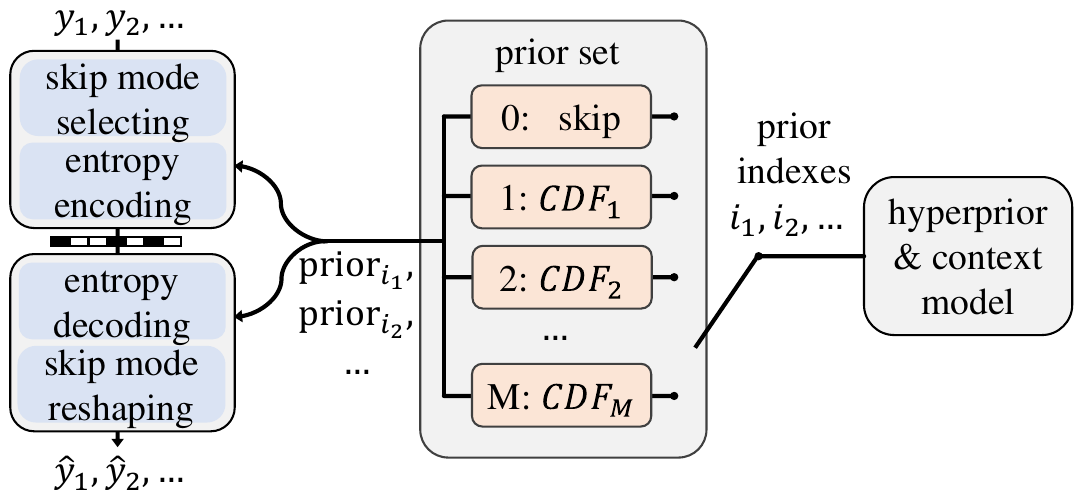}
    \caption{Diagram of coding process of the proposed switchable priors method. The Cumulative Distribution Function is abbreviated as CDF. The prior set contains various prior distributions described by a skip mode and CDF tables shared by both the encoder and decoder. The entropy model predicts the appropriate prior index within the set for each latent, rather than estimating the specific parameters of parameterized probabilistic models. The encoder selects latents based on skip mode and entropy encodes the selected latents using the prior set and predicted indexes. The decoder recovers the coded latents from bitstreams and then reshapes them to the original position.
    }
    \label{fig:diagram_switch}
\end{figure}

Entropy coding in neural image compression relies on a prior distribution of discrete latents, commonly referred to as the entropy model, which is shared between the encoder and decoder.
Advanced neural image compression methods usually employ conditional entropy models to characterize the distribution of latent variables, such as \cite{minnen2020channel, he2022elic, mentzer2023m2t, li2023flexible, fu2023learned, jiang2023mlic,li2023frequency, li2024mixer, han2024cca, zhang2024decouple, guo2021causal, li2024mpem, ge2024nlic, koyuncu2024contextformer}. These models adopt a predefined family of probabilistic distributions as the prior of the latent variables, such as the Gaussian probabilistic model. The parameterized prior distribution is conditioned on parameters derived from additional neural networks, such as the hyperprior module \cite{balle2018variational} and context models \cite{minnen2018joint}.
For example, in \cite{balle2018variational}, a hyperprior was employed to predict the scale parameters of the Gaussian model for describing the distribution of latents. 

Probabilistic models with more complex forms can more effectively characterize the distribution of latents in neural image compression, thereby reducing the rate and improving compression performance. Some studies have explored mixture models as the probabilistic model, such as the Gaussian mixture model \cite{cheng2020learned}. Zhang \textit{et. al.} \cite{zhang2024ggm} introduced the generalized Gaussian model which achieves better compression performance. Compared to the Gaussian model, these probabilistic models with more complex forms effectively improve performance, while also bringing higher complexity, limiting their practical usage. Specifically, the computational cost for network inference increases due to the larger number of parameters that need to be predicted. Moreover, complex probabilistic models incur additional computational and memory costs for constructing the Cumulative Distribution Function (CDF) tables, which are necessary for entropy coders such as arithmetic coder \cite{langdon1984ae} and asymmetric numeral systems \cite{duda2013asymmetric}.

In this paper, to enhance the practical usability of advanced probabilistic models such as the generalized Gaussian model \cite{zhang2024ggm}, we propose learning switchable priors to decouple entropy model complexity from the adopted probabilistic model. Specifically, we construct a finite set of trainable priors containing a variety of parameterized probability distributions. Each prior in this set is defined by a group of trainable parameters and a parametric probabilistic model. Instead of directly estimating specific parameters of the adopted probabilistic model, we train the entropy model to predict the index of the appropriate prior in this set. During training, the parameters of each distribution in the prior set, along with those hyperprior or context models, are jointly optimized.
After training, each distribution is converted into a CDF table, which is pre-stored and shared by the encoder and decoder for entropy coding. During testing, entropy coders utilize these pre-stored CDF tables and the indexes predicted from the entropy model to encode and decode discrete latents. This approach enables the adoption of more complex probabilistic models without increasing the complexity of actual coding.
Moreover, compared to dynamically constructing CDF tables during testing, our method eliminates the associated computational and memory costs. Compared to look-up tables-based methods \cite{balle2019integer}, which also pre-store CDF tables but require dynamically mapping predicted parameters to indexes, our approach requires fewer CDF tables and avoids the computational cost of building indexes.

To further enhance compression performance and reduce complexity, we introduce several improvements to the basic switchable priors idea. 
First, we incorporate a skip mode into the prior set, enabling the model to omit latents that negatively influence performance, which could accelerate the entropy coding process and improve the compression performance. The diagram of the coding process of the proposed switchable priors method with the skip mode is shown in Fig. \ref{fig:diagram_switch}. Second, we reuse the prior set of the main latents for hyperlatents, which further reduces the storage cost of CDF tables for coding hyperlatents without affecting compression performance. Third, to accelerate rate estimation during training switchable priors, we propose selecting the Top-K most likely distributions from the prior set rather than calculating the weighted rate across all possible distributions. 

To provide a practical and efficient solution for neural image compression, we further present the FastNIC model along with the proposed switchable priors approach. Many existing neural image compression methods suffer from high computational complexity, which limits their applicability in real-world scenarios. To address this challenge, we designed FastNIC to significantly reduce both computational complexity and coding time while maintaining better compression performance than BPG when trained for mean squared error (MSE). Specifically, among the NIC methods that outperform BPG, FastNIC achieves the to-date lowest computational complexity and the fastest coding speed, demonstrating its potential for practical deployment. Furthermore, when trained for multi-scale structural similarity (MS-SSIM), FastNIC exhibits significantly better compression performance than VTM, highlighting its practical value when perceptual quality is concerned.

In addition to FastNIC, we also validate the effectiveness of the switchable priors method on other advanced models, including the Shallow decoder \cite{yang2023shallow}, ELIC \cite{he2022elic}, TCM \cite{liu2023learned}, and FM-intra model \cite{li2024fm}. Experimental results demonstrate that the proposed switchable priors method effectively reduces entropy coding complexity while improving compression performance.

Our main contributions can be summarized as follows:
\begin{itemize}
    \item We propose learning switchable priors to decouple the complexity of actual coding from the form of the adopted probabilistic model. This approach enables the use of more advanced probabilistic models to enhance compression performance without increasing complexity.
    \item We present FastNIC, a practical and efficient solution for neural image compression. By incorporating switchable priors, it outperforms BPG on the Kodak dataset and achieves encoding and decoding complexities under 12 KMACs/pixel and 10 KMACs/pixel, with both encoding and decoding times around 100 ms for a 4K image.
    \item Experimental results show that the proposed switchable priors method improves compression performance and significantly reduces entropy coding complexity.
\end{itemize}

\section{Related Work}
\subsection{Neural Image Compression}
In the typical scheme of neural image compression \cite{balle2017end}, the encoder applies an analysis transform to an input image, producing latent representations. These latents are then quantized and losslessly encoded into bitstreams using an entropy model. Finally, the decoder recovers the quantized latents and reconstructs the image through the synthesis transform. 

Non-linear transforms are crucial for neural image compression.
Recent studies have incorporated more advanced architectures, such as residual connections and attention mechanisms, to enhance compression performance \cite{cheng2020learned, guo2021causal, he2022elic, liu2023learned, li2023frequency, ge2024nlic}. 
Moreover, some studies have proposed using invertible transforms to facilitate compression \cite{ma2020end, dong2024wavelet}.

Entropy models are crucial for estimating the prior distribution of latent variables. A factorized prior was first introduced in \cite{balle2017end}. Subsequently, a hyperprior model was proposed in \cite{balle2018variational}, which parameterizes the distribution of latent variables as a Gaussian model conditioned on side information. Minnen \textit{et al.} \cite{minnen2018joint} further improved the entropy model by jointly leveraging a context model and hyperprior to achieve higher accuracy. Numerous studies \cite{minnen2020channel, he2022elic, mentzer2023m2t, li2023flexible, fu2023learned, jiang2023mlic,li2023frequency, li2024mixer, han2024cca, zhang2024decouple, guo2021causal, li2024mpem, ge2024nlic, koyuncu2024contextformer} have focused on advancing entropy models to enhance performance.

The derivative of the quantization operator is zero almost everywhere, rendering standard back-propagation inapplicable during training. Various quantization surrogates have been proposed to enable end-to-end training. 
Minnen and Singh \cite{minnen2020channel} empirically proposed a mixed quantization surrogate that uses noisy latents for rate estimation but employs rounded latents with the straight-through estimator \cite{bengio2013estimating} when passing through the synthesis transform. This mixed quantization surrogate has been widely used in recent studies.

\subsection{Computationally Efficient Neural Image Compression}
For practical applications, several studies have focused on constructing computationally efficient neural image compression methods.
Some studies \cite{minnen2023advancing, zhang2023practical} empirically proposed more efficient network structures for neural image compression. Other studies \cite{ yang2021slimmable, luo2022slimhyper, wang2023evc} employed pruning and distillation techniques to derive computationally efficient transforms. Yang \textit{et al.} \cite{yang2023shallow} proposed an asymmetrical architecture with a powerful encoder and a shallow decoder to reduce decoding complexity. 
For the parallel computation of spatial context models, He \textit{et al.} \cite{he2021checkerboard} introduced a checkerboard context model, while Minnen and Singh \cite{minnen2020channel} proposed a fast channel-wise autoregressive model. 
To accelerate entropy coding, Shi \textit{et al.} \cite{shi2022alphavc} proposed skipping latent variables whose scale parameters fall below a predefined threshold, while Lee \textit{et al.} \cite{lee2022selective} introduced joint training with additional networks to predict a binary mask indicating latents to skip. The details of the skipping mechanism are included in Sec. \uppercase\expandafter{\romannumeral1} in the supplementary material.

\subsection{Probabilistic Models for Neural Image Compression}
The probabilistic model plays a crucial role in characterizing the distribution of latent variables in neural image compression. Probabilistic models with more parameters and complex forms can more accurately represent the distribution, thereby enhancing compression performance.
However, due to complexity constraints, conditional entropy models in neural image compression often adopt parametric probabilistic models with fewer parameters.
In \cite{balle2018variational}, a zero-mean Gaussian scale model is adopted to estimate the distribution of latent variables, where a hyperprior module is used for predicting scale parameters. In this scheme, the Cumulative Distribution Function (CDF), essential for entropy coding, must be constructed dynamically during encoding and decoding. Minnen \textit{et al.} \cite{minnen2018joint} extended this approach to the mean-scale Gaussian model. More advanced probabilistic models, such as the Gaussian Mixture Model (GMM) \cite{cheng2020learned}, and the Gaussian-Laplacian-Logistic mixture model \cite{fu2023learned}, were introduced to improve distribution modeling capabilities, achieving better compression performance. Zhang \textit{et al.} \cite{zhang2024ggm} presented the generalized Gaussian model (GGM), which introduces only one additional shape parameter compared to the Gaussian model and outperforms both the Gaussian model and GMM on various neural image compression models. Despite their performance advantages, these more complex probabilistic models bring higher complexity compared to the Gaussian model, thereby limiting their practical usage.

\subsection{Entropy Coding for Neural Image Compression}
Arithmetic coders \cite{langdon1984ae} and asymmetric numeral systems \cite{duda2013asymmetric} are widely employed in neural image compression methods to encode discrete latents into bitstreams. During encoding, the entropy encoder requires both the latents $\hat{y}$ and the corresponding CDF tables as inputs to encode the symbols into bitstreams. In the decoding process, the same CDF tables must be provided to the entropy decoder to decompress $\hat{y}$ correctly. For conditional entropy models, which utilize hyperprior and context models, CDF tables for latent variables must be constructed dynamically during encoding and decoding. This dynamic construction introduces additional computational costs, high memory usage, and potential floating-point errors. 

To address these issues, most recent studies \cite{liu2023learned,li2024fm,jiang2023mlic,begaint2020compressai} adopted a method based on look-up tables (LUTs) \cite{balle2019integer}. The LUTs-based method only influences the test stage and does not influence the training process. 
After training, several entropy parameter values of the probabilistic model are sampled in a specific manner from their respective possible ranges. Then, the corresponding CDF table is calculated for each sampled value. The encoder and decoder share these pre-computed CDF tables. During the actual encoding and decoding process, the predicted entropy parameters, which model the distribution of latent variables, are quantized to the nearest sampled value through a searching process. These quantized values are then used to index the corresponding CDF table, which is necessary for entropy coders.
The LUTs-based approach eliminates the need for dynamic calculation of CDF tables during encoding and decoding, thereby saving computational costs and reducing coding time.
Ballé \textit{et al.} \cite{balle2019integer} first adopted this method for the zero-mean Gaussian scale model.
A simpler way to build indexes for the Gaussian scale model was employed in the reference software of JPEG-AI \footnote{Available online at \url{https://gitlab.com/wg1/jpeg-ai/jpeg-ai-reference-software}.}.
Sun \textit{et al.} \cite{sun2021learned} extended the LUTs-based implementation for the mean-scale Gaussian model by utilizing two-dimensional CDF tables. He \textit{et al.} \cite{he2022post} attempted to use the LUTs-based approach for GMM, where CDF tables are shared across different Gaussian components, but the weighted summation of each component's CDF is still computed dynamically. Zhang \textit{et al.} \cite{zhang2024ggm} extended this method to GGM, where two-dimensional CDF tables are established considering two parameters for indexing the corresponding CDF. However, extensions of the LUTs-based method for more complex probabilistic models increase the cost required for storing CDF tables and the computational cost of building indexes compared to the Gaussian model.

\begin{figure}[!t]
  \centering
\includegraphics[width=0.74\linewidth]{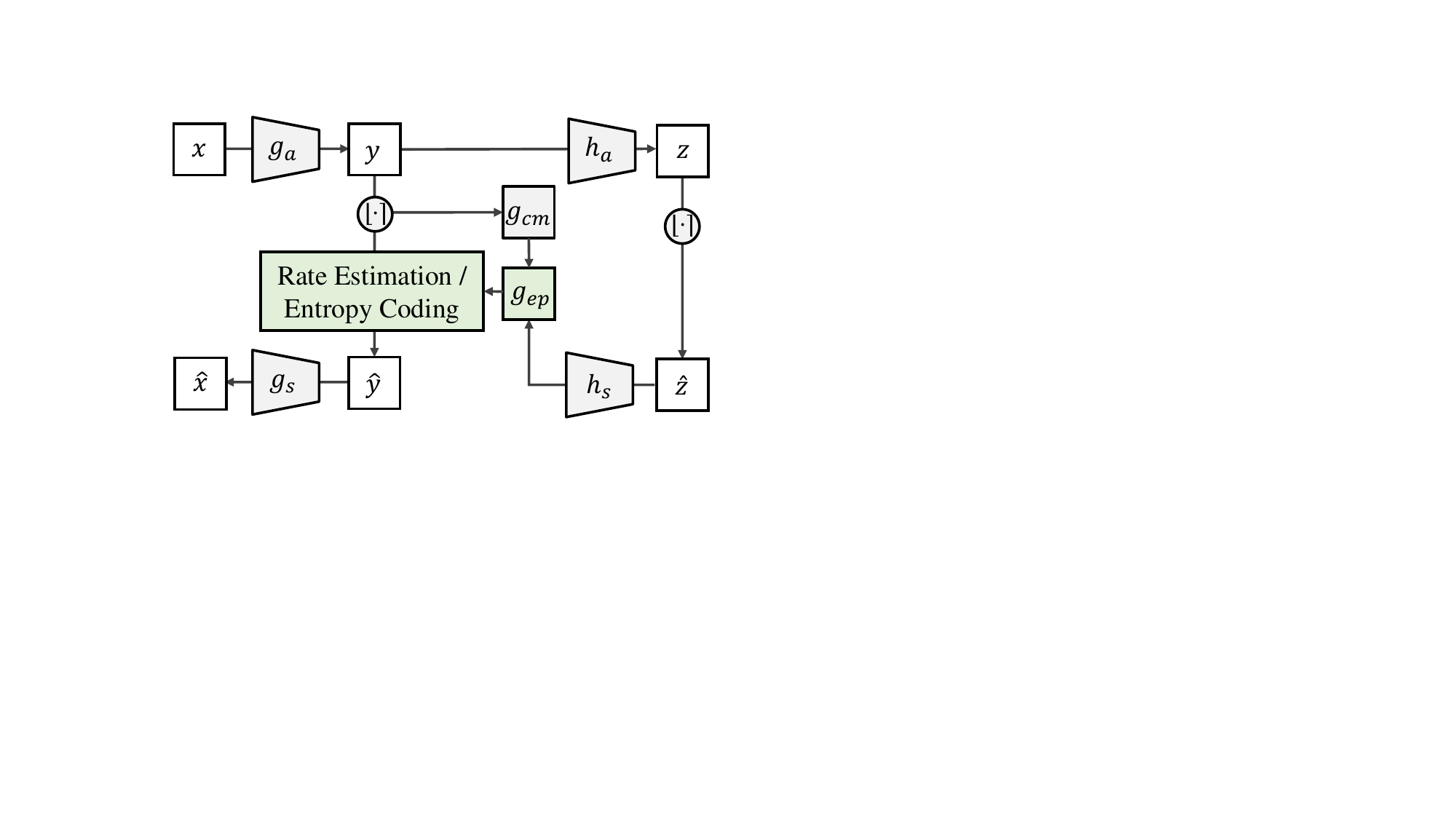}
    \caption{Diagram of a typical neural image compression model. $g_a$ and $g_s$ denote the analysis and synthesis transforms. $h_a$ and $h_s$ denote the hyper-analysis and synthesis transform. $g_{cm}$ denotes the context model. $g_{ep}$ denotes the entropy parameters module, which is used to generate the entropy parameters of probabilistic models. Our switchable priors method focuses on the improvement in $g_{ep}$, the rate estimation process during training, and the entropy coding during testing.
    }
    \label{fig:diagram_e2e}
\end{figure}

\section{Learning Switchable Priors}
The core concept of the switchable priors method is presented in Sec. \ref{sec:switch}. The extension for learning a multi-dimensional prior set is introduced in Sec. \ref{sec:multidim}. The incorporation of the skip mode is discussed in Sec. \ref{sec:skippping}. Reusing the prior set for hyperlatents is detailed in Sec. \ref{sec:jointyz}.

\subsection{Neural Image Compression with Switchable Priors}
\label{sec:switch}
In the typical neural image compression model shown in Fig. \ref{fig:diagram_e2e}, the prior distributions of latents are conditioned on $\theta$, where $\theta$ is the continuous parameter of the adopted probabilistic model obtained from the $g_{ep}$.
During training, with the assumption that the CDF of the adopted probabilistic model is $F(\cdot|\theta)$, the rate of $\hat{y}$ estimated by the entropy model is
\begin{equation}
\begin{split}
R&(\hat{y}) = -\log_2 P(\hat{y}|\theta),\\
\mbox{where }P(\hat{y}|\theta)&=F(\hat{y}+0.5|\theta)-F(\hat{y}-0.5|\theta).
\end{split}
\label{eq:rate_estimation}
\end{equation} 

\begin{figure*}[!t]
  \centering
    \includegraphics[width=0.84\linewidth]{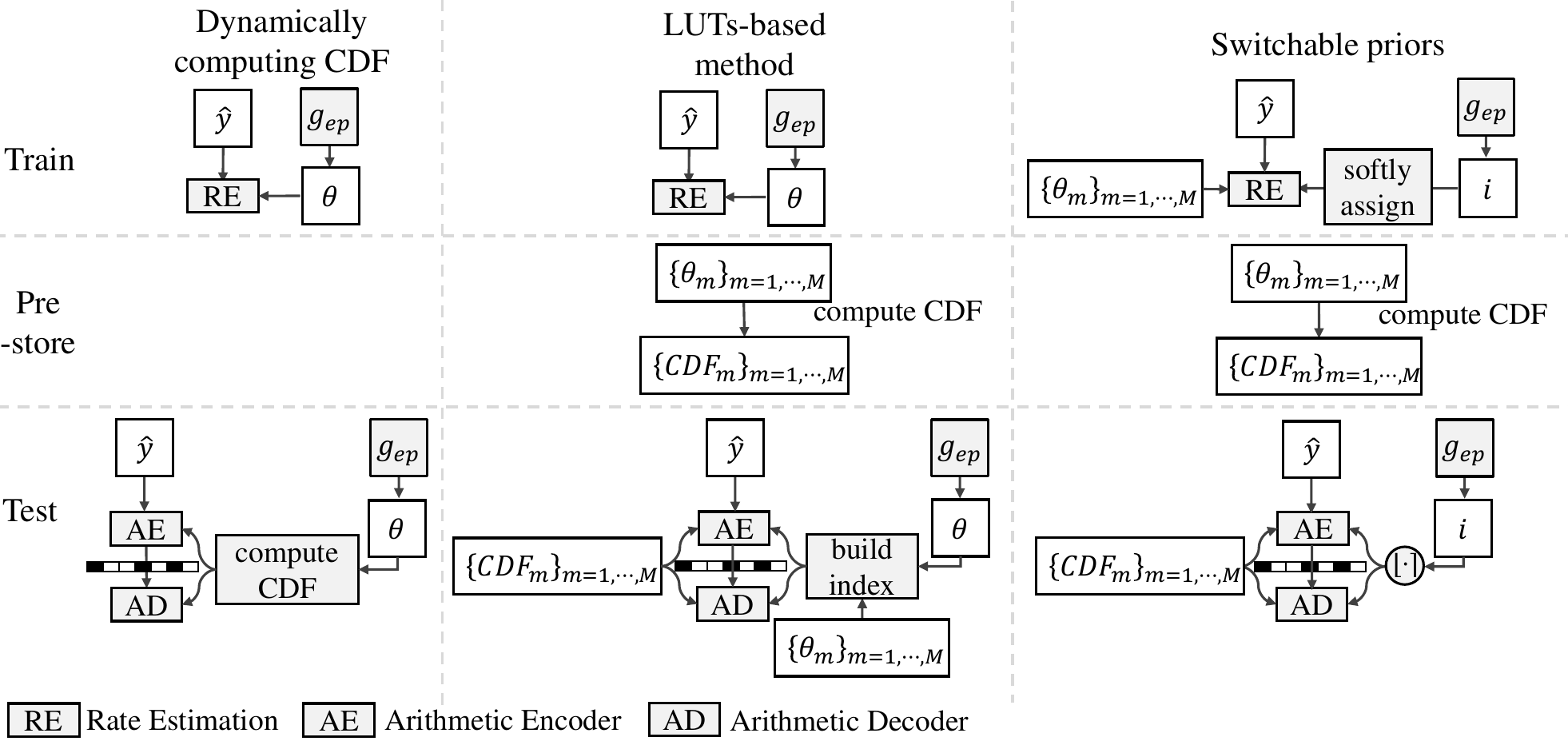}
    %\vspace{-1em}
    \caption{Diagram of training and testing stages for dynamically computing CDF, LUTs-based method, and the proposed switchable priors method.}
\label{fig:diagram_three_entropy_coding}
\end{figure*}

At the test stage, as shown in Fig. \ref{fig:diagram_three_entropy_coding}, entropy coders require the CDF table of each element in $\hat{y}$ to conduct entropy coding. The CDF tables must be dynamically constructed based on $\theta$ predicted by the entropy model. The number of CDF tables equals the number of latents that need to be coded. This dynamic construction process consumes a lot of time due to the additional computational costs and memory access time required for substantial CDF tables.
To address these issues, most recent studies \cite{begaint2020compressai,liu2023learned,li2024fm} adopted the method \cite{balle2019integer} based on look-up tables (LUTs). As shown in Fig. \ref{fig:diagram_three_entropy_coding}, the LUTs-based approach only influences the test stage and does not influence the training process. After training, several $\theta$ values $\{\theta_m\}_{m=1,\cdots,M}$ are sampled in a specific way from the possible range of $\theta$. Then, the corresponding CDF table is calculated for each $\theta_m$. These pre-computed CDF tables are shared by the encoder and decoder for entropy coding. During testing, the predicted $\theta$ is quantized to the closest $\theta_m$ in $\{\theta_m\}_{m=1,\cdots,M}$, and then used to index the corresponding CDF table. Since the number of required CDF tables, $M$, is typically much smaller than the number of coded latents, the LUTs-based approach significantly reduces memory costs and memory access time. Additionally, the LUTs-based approach eliminates the dynamical construction process at the actual coding process, thereby reducing computational costs.

The entropy parameter $\theta$ can comprise multiple parameters, depending on the probabilistic model used. For example, when applying zero-center quantization, \textit{i.e.} coding $\lfloor y-\mu \rceil$, $\theta = \sigma$ in the Gaussian model \cite{balle2020nonlinear}, where $\mu$ and $\sigma$ denote the mean and scale parameters, respectively, and $\theta = \{\alpha, \beta\}$ in the generalized Gaussian model (GGM) \cite{zhang2024ggm}, where $\alpha$ and $\beta$ denote the scale and shape parameters, respectively.
GGM introduces the additional shape parameter $\beta$ compared to the Gaussian and it degenerates into Gaussian when $\beta=2$ (with $\sigma=\alpha/\sqrt{2}$). 
For a Gaussian mixture model with three Gaussian components, which typically applies nonzero-center quantization, \textit{i.e.} coding $\lfloor y \rceil$, $\theta = \{p_1, \mu_1, \sigma_1, \dots, p_3, \mu_3, \sigma_3\}$, where $p_i$, $\mu_i$, and $\sigma_i$ denote the weight, mean, and scale parameters of the $i$-th Gaussian component, respectively. 

Probabilistic models with more complex forms can more effectively characterize the distribution of latents, thereby improving compression performance. 
However, existing methods for entropy coding, such as dynamically computing CDF tables or LUTs-based approaches, face challenges when incorporating more complex probabilistic models. 
Specifically, the network complexity increases due to the larger number of parameters that need to be predicted. Moreover, dynamically constructing CDF tables with complex probabilistic models incurs additional computational costs.
For the LUTs-based method, probabilistic models with more parameters require finer parameter sampling levels, consuming more CDF tables for entropy coding. This causes higher costs for storing CDF tables, longer memory access time for entropy coding, and higher computational costs for building indexes of corresponding CDF tables.
These limitations hinder the practical adoption of more advanced probabilistic models.

To address these challenges, we propose learning switchable priors to decouple the complexity of neural image compression models from the adopted probabilistic models, as illustrated in Fig. \ref{fig:diagram_three_entropy_coding}.
Specifically, we construct a finite trainable prior set containing various parameterized probability distributions, denoted as $\{\theta_m\}_{m=1,\cdots,M}$, where $M$ is the number of distributions in the prior set. Each prior distribution in this set is defined by a group of trainable parameters and the corresponding probabilistic model.
Instead of directly estimating the parameters of distributions, the entropy model predicts the index of the appropriate prior within this set for each element in $\hat{y}$. The training, pre-storing, and testing of the switchable priors method are introduced in the following. 

\textbf{Training.}
During training, the parameters of each prior distribution and the entropy model are jointly optimized. The index obtained from $g_{ep}$ is a continuous value, denoted as $i$. Rate estimation is essential for jointly optimizing the parameters of each prior distribution along with the entropy model. 
To enable the training of each prior, the rate is estimated through a weighted sum of the rate calculated by each prior, as
\begin{equation}
R(\hat{y}) = \sum_{m=1}^{M} -\pi_m \log_2 P(\hat{y}|\theta_{m}),
\label{eq:weight_rate}
\end{equation}
where $\pi_m$ represents the weight assigned to the $m$-th prior, $P(\hat{y}|\theta_{m})$ is the probability estimated by the $m$-th prior, and $M$ is the number of priors. 

To train the entropy model to predict the prior index, we establish a connection between the predicted index value $i$ obtained from $g_{ep}$ and the weight of each prior $\pi_m$. Specifically, we use the negative distance, $-|i-m|$, to weigh each prior. Then, we use the softmax operator to obtain the weight of each prior $\pi_m$. The weight of the $m$-th prior is given by
\begin{align}
\pi_m = \frac{\exp\left(-\frac{|i-m|}{\tau}\right)}{\sum_{m=1}^{M}\exp\left(-\frac{|i-m|}{\tau}\right)},
\label{eq:soft_assign}
\end{align}
where $\tau>0$ is a temperature parameter. Indeed, the commonly used softmax operation ensures that the weights satisfy $\sum_{m=1}^{M} \pi_m = 1$. This design also ensures that the prior whose index is closest to $i$ will have the largest weight.

During training, as $\tau$ approaches 0, the weights for priors tend towards a one-hot form, meaning each symbol is effectively assigned to a single prior distribution from the set.

\textbf{Pre-storing.}
After training, $\{\theta_m\}_{m=1,\cdots,M}$ are converted to $\{CDF_m\}_{m=1,\cdots,M}$ by computing the CDF table of each prior. The CDF tables are pre-stored and shared by the encoder and decoder for entropy coding.

\textbf{Inference.}
At the inference stage, the index of the optimal prior distribution is determined by maximizing $\pi_m$, which can be simplified by 
\begin{equation}
\begin{split}
\text{index} &= \mathop{\arg\max}\limits_{m\in \{1,2,\cdots,M\}} \pi_m \\
&= \mathop{\arg\max}\limits_{m\in \{1,2,\cdots,M\}} -|i-m|\\
&= \mathop{\arg\min}\limits_{m\in \{1,2,\cdots,M\}} |i-m|\\
&=  \lfloor\text{Clip}( i ,1,M)\rceil,
\end{split}
\label{eq:build_index}
\end{equation}
where $\text{Clip}(i,1,M)$ means clipping $i$ into $[1,M]$ and $\lfloor\cdot\rceil$ denotes the rounding operator.
Therefore, we only need to discretize the continuous index value predicted by the entropy model to obtain the index of the CDF table for entropy coding. This process efficiently maps the predicted $i$ into the prior set.

The switchable priors method ensures that the complexity of actual coding remains independent of the adopted probabilistic model. This allows us to incorporate more advanced probabilistic models such as the generalized Gaussian model \cite{zhang2024ggm} without increasing complexity. 
Compared to dynamically constructing CDF tables for entropy coding, our method eliminates the extra computational and memory costs. In contrast to the LUTs-based method, our approach could reduce the number of CDF tables required for entropy coding and avoid the computational cost of building indexes required for more complex probabilistic models.

\subsection{Learning Multi-Dimensional Prior Set}
\label{sec:multidim}
For the previously discussed switchable priors, the goal is to identify the appropriate distribution by predicting a one-dimensional index. This approach relies on effectively projecting the parameters of the probabilistic model into a one-dimensional space. However, this can be challenging for probabilistic models with very high degrees of freedom.

To address this limitation and improve the generalization of switchable priors, we propose an extension that expands both the prior set and the predicted indexes to higher dimensions. This extended framework allows for the use of significantly more complex probabilistic models when needed. Importantly, the use of a multi-dimensional prior set is not mandatory, and a one-dimensional prior set may suffice if it provides adequate performance.

We show an extension for a two-dimensional prior set, defined by $\{\theta_{m,n}\}_{m=1,\cdots,M;n=1,\cdots,N}$. The predicted two-dimensional continuous index values from $g_{ep}$ are $(i,j)$. The weight of each prior distribution can be formulated by 
\begin{equation}
\begin{split}
\pi_{m,n}&=\frac{\exp(-\frac{|i-m|}{\tau})}{\sum_{m=1}^{M}\exp(-\frac{|i-m|}{\tau})}\frac{\exp(-\frac{|j-n|}{\tau})}{\sum_{n=1}^{N}\exp(-\frac{|j-n|}{\tau})}.
\end{split}
\label{eq:multi_dim}
\end{equation}
After training, the set is converted to two-dimensional CDF tables. The index built at the test stage is $\left(\lfloor\text{Clip}( i ,1,M)\rceil, \lfloor\text{Clip}( j ,1,N)\rceil\right)$.

\subsection{Top-K Acceleration for Training}
During training, the rate estimation for the switchable priors method involves calculating the probabilities for all distributions in the prior set, which can be computationally expensive if the prior set is large. To accelerate the training process, we propose a Top-K acceleration strategy. Specifically, for rate estimation, we select only the top $K$ distributions with the largest weights. The following shows how the Top-K acceleration strategy is applied to the one-dimensional prior set. From Eq. (\ref{eq:build_index}), we have
$\arg\max_{m} \pi_m= \arg\min_{m} |i-m|$.
Therefore, we only need to find the nearest $K$ values to $i$ from $\{1,2,\cdots,M\}$, as given by
\begin{align}
\label{eq:top_k_2}
    \{m_1,m_2,\cdots,m_K\} = \mathop{\arg\text{Min-K}}\limits_{m\in \{1,2,\cdots,M\}}|i-m|.
\end{align}
Then, the estimated rate is given by
\begin{equation}
\begin{split}
&R(\hat{y}) = \sum_{k=1}^{K} -\pi_{m_k}\log_2 P(\hat{y}|\theta_{m_k}),\\
&\mbox{ where }\pi_{m_k} = \frac{\exp(-\frac{|i-m_k|}{\tau})}{\sum_{k=1}^{K}\exp(-\frac{|i-m_k|}{\tau})}.
\end{split}
\label{eq:top_k_rate}
\end{equation}

\textbf{Top-2 acceleration.} We can simplify the process by selecting two samples to calculate the estimated rate for efficient training. In this case, the calculation of the nearest 2 values to $i$ can be expressed as  
\begin{align}
\label{eq:top_2}
    m_1=\lfloor \text{Clip}(i,1,M) \rfloor, m_2=\lceil \text{Clip}(i,1,M) \rceil.
\end{align}
We employ Top-2 acceleration for efficient training in our method. When learning a multi-dimensional prior set, the Top-2 acceleration is applied separately to each dimension.

\subsection{Incorporation with Skip Mode}
\label{sec:skippping}

\begin{figure}[!t]
  \centering
  \subfloat[Avg. bpp: 0.11 ($\lambda=0.0018$)]
    {  
        \includegraphics[width=0.48\linewidth]{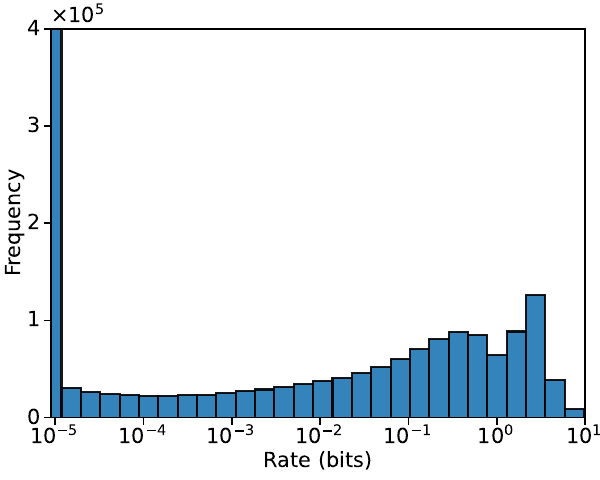}
        \label{fig:fre_0.0018}
    }\hspace{-0.3cm}
    \subfloat[Avg. bpp: 0.85 ($\lambda=0.0483$)]
    {  
        \includegraphics[width=0.48\linewidth]{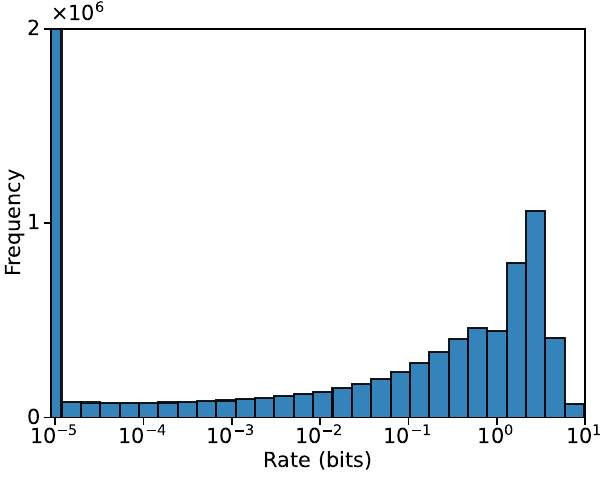}
        \label{fig:fre_0.0483}
    } 
    \caption{Frequency of the rate of latent variables collected from the Kodak set in the FM-intra model \cite{li2024fm}. The x-axis is log-scaled.
    }
    \label{fig:frequency_FM}
\end{figure}

In neural image compression models, a significant portion of coded latents exhibits extremely minimal rates, as shown in Fig. \ref{fig:frequency_FM}. Although these latents contribute negligibly to the overall rate, they still require corresponding priors for entropy coding within the switchable priors framework, which unnecessarily increases storage and computational costs. To mitigate this, we introduce a skip mode as an additional prior in the switchable priors framework. The skip mode allows for bypassing entropy coding for latents with negligible rates, thus reducing the number of CDF tables required. Additionally, by omitting the entropy coding process for partial latents, the skip mode also reduces computational complexity, providing a practical improvement.

For the skip mode, a binary mask $\hat{b}$ is introduced to indicate whether each element is skipped or reserved during entropy coding. Specifically, $0$ denotes a skipped latent, while $1$ indicates a reserved one.
Similar to \cite{lee2022selective}, our skipping mechanism also utilizes additional networks to predict the binary mask. Specifically, we use $g_{ep}$ to produce $b$, which is then employed to calculate the logits for the binary mask. Additionally, we have developed more effective training methods to enhance performance. 

First, we optimize the additional network using the true rate-distortion objective estimated by rounded latents rather than the proxy objective caused by quantization approximation during training in \cite{lee2022selective}. Specifically, we separately train the additional network based on a well-trained neural image compression model. During the optimization of the additional network, all other modules in the neural image compression model are kept fixed. This design allows us to use rounded latent variables to estimate the rate-distortion cost, thereby optimizing the true objective.
The objective during training is given by
\begin{align}
L=R(\hat{y})\cdot\tilde{b}+R(\hat{z})+\lambda D(x,g_s(\hat{y}\cdot\tilde{b})),
\label{eq:mask_loss}
\end{align}
where $\lambda$ controls the trade-off between rate and distortion, and $\tilde{b}$ is the approximate binary mask used during training.

Second, to enable end-to-end training of skip mode, we employ the Gumbel-Softmax \cite{jang2017gumbel} estimator to enable gradient propagation through the binary mask generation process. This method provides a more accurate gradient estimation compared to the simple straight-through estimator \cite{bengio2013estimating} in \cite{lee2022selective}. The mask generation process during training can be formulated as 
\begin{equation}
\begin{split}
    &\text{logit}_0 = -\frac{|b-0|}{t},\text{logit}_1 =  -\frac{|b-1|}{t},\\
    &\tilde{b} = \text{Gumbel-Softmax}(\text{logit}_0,\text{logit}_1),
\end{split}
\label{eq:mask}
\end{equation}
where logit value indicates the confidence score of mask 0 and 1, $t$ is a temperature parameter, and $\tilde{b}$ is the approximate sample for the binary mask through Gumbel-Softmax. The Gumbel-Softmax operator approximates the discrete sampling process with a continuous relaxation, enabling back-propagation through the mask generation. During training, with the temperature parameter approaching 0, $\tilde{b}$ gradually becomes a hard binary mask.

At the test stage, $b$ is discretized to get a hard binary mask 
$\hat{b} = \lfloor \text{Clip}(b,0,1) \rceil$.

\subsection{Reusing Prior Set for Hyperlatents}
\label{sec:jointyz}
Typically, the entropy model for $\hat{z}$ employs a factorized prior, where a unique CDF table is assigned to each channel of $\hat{z}$. As a result, the total number of required CDF tables equals the number of channels in $\hat{z}$.
The switchable priors method can also be applied to the coding for hyperlatents $\hat{z}$. This extension could further reduce the cost of the CDF tables associated with $\hat{z}$ while maintaining performance.

Specifically, we reuse the same prior set of $\hat{y}$ to code $\hat{z}$. We use the one-dimensional prior set to show this reusing strategy.
During training, we introduce additional trainable logits $l_1,\cdots,l_M$ to predict the corresponding prior index for each channel in $\hat{z}$. The estimated rate for $\hat{z}$ is given by
\begin{equation}
\begin{split}
&R(\hat{z}) = \sum_{m=1}^{M} -\pi_{z_m}\log_2 P(\hat{z}|\theta_m),\\
&\mbox{where }\pi_{z_m} = \frac{\exp(l_m)}{\sum_{m=1}^{M}\exp(l_m)}.
\end{split}
\label{eq:hyper}
\end{equation}
These logits are jointly trained with the prior set. Since the number of elements in $\hat{z}$ is relatively small, we do not apply the Top-K acceleration strategy for calculating $R(\hat{z})$. 
After training, the index of the optimal prior for coding $\hat{z}$ is fixed as $\arg\max_{m} {\pi_{z_m}}$.

For skip mode, we introduce one additional trainable $b_z$ and generate the mask $\tilde{b}_z$ as depicted in Eq. (\ref{eq:mask}) during training.
The training objective is given by
\begin{align}
L = R(\hat{y})\cdot\tilde{b}+R(\hat{z})\cdot\tilde{b}_z+\lambda D(x,g_s(\hat{y}\cdot\tilde{b})).
\label{eq:hypermask}
\end{align}
After training, $b_z$ is discretized to obtain a binary mask $\hat{b}_z = \lfloor Clip(b_z,0,1) \rceil$, and channels indicated by 0 are pruned.

\section{FastNIC: A Lightweight Neural Image Compression Model}
\label{sec:fastnic}
To provide a practical and efficient solution for neural image compression, we further present the FastNIC model 
and evaluate the practicality of the switchable priors method on the FastNIC model. Many existing neural image compression methods suffer from high computational complexity, which limits their applicability in real-world scenarios. To address this challenge, we design FastNIC to significantly reduce both computational complexity and coding time.

\begin{figure}[!t]
  \centering
  \subfloat[Diagram.]
    {  \includegraphics[width=0.7\linewidth]{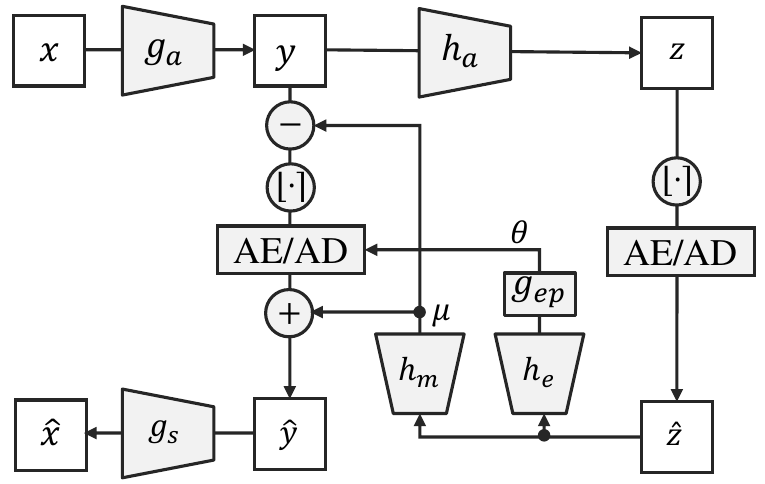}
        \label{fig:diagram_fastnic}
    }
    
    \subfloat[Network structure.]
    {  
        \includegraphics[width=0.98\linewidth]{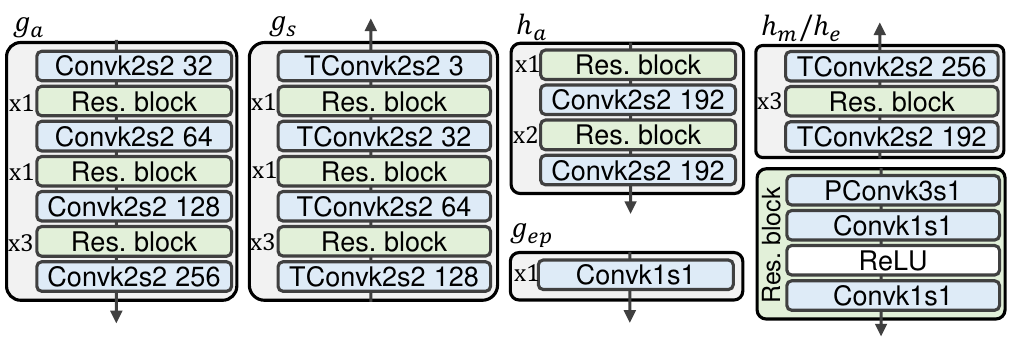}
        \label{fig:network_fastnic}
    } 
    \caption{Framework of the proposed \textbf{FastNIC} model. (a) shows the diagram of FastNIC. In the entropy model, the module $h_m(\text{mean})$ is used to predict the mean parameter, while the module $h_e(\text{entropy})$ predicts other parameters of the probabilistic models. The detailed network structure is shown in (b).
    }
    \label{fig:fastnic_structure}
\end{figure}

We employ the hyperprior structure to construct the FastNIC model, as shown in Fig. \ref{fig:diagram_fastnic}. The main compression process can be formulated as 
\begin{align}
y = g_a(y),~\hat{y} = \lfloor y-\mu\rceil+\mu,~\hat{x}=g_s(\hat{y}),
\label{eq:main_loop}
\end{align}
where $\mu$ is the mean parameter of probabilistic models such as the Gaussian model and the generalized Gaussian model. The symbols coded into bitstreams are $\lfloor y-\mu \rceil$.
The prior distribution of $\hat{y}$ is obtained through
\begin{align}
z = h_a(y),~\hat{z} = \lfloor z\rceil,~\mu=h_m(\hat{z}),~\theta=g_{ep}(h_e(\hat{z})),
\label{eq:hyper_loop}
\end{align}
where $\mu$ and $\theta$ are parameters of probabilistic models. For example, $\theta=\sigma$ for the Gaussian model and $\theta=\{\beta,\alpha\}$ for the generalized Gaussian model.

For applying the proposed switchable priors method, we simply adjust the output of $g_{ep}$ as $i,b=g_{ep}(h_e(\hat{z}))$,
where $b$ is used to indicate the mask map for skipping and $i$ is used to indicate the index of distribution in the prior set.

To reduce computational costs, we focus on designing more efficient transform modules, as illustrated in Fig. \ref{fig:network_fastnic}. First, we adopt a single convolution layer with a stride of 2 for both down-sampling and up-sampling operations. To further reduce complexity, the kernel size in these layers is reduced to 2, as suggested in \cite{zhang2023practical}. Second, residual blocks are employed to improve the capacity of the transforms, following \cite{he2022elic}. Specifically, we utilize the advanced FasterNet block \cite{chen2023pconv}, which incorporates partial convolution to balance complexity and performance. The partial convolution layer processes spatial dense convolution only on partial channels. Third, we gradually widen the channel of features and concentrate more parameters at lower-resolution features, instead of using an equivalent number of channels and parameters across different feature resolution stages \cite{balle2018variational, cheng2020learned, he2022elic}. This design effectively reduces overall computational complexity. Our FastNIC achieves encoding complexity below $12$ KMACs/pixel and decoding complexity below $10$ KMACs/pixel. 

\section{Experiments}
\subsection{Experimental Setting}
\subsubsection{Models}
We validate the effectiveness of the switchable priors method on several neural image compression methods:
\begin{itemize}
    \item \textbf{FastNIC} represents our proposed lightweight neural image compression model specified in Sec. \ref{sec:fastnic}.
    \item \textbf{ELIC} (CVPR2022) denotes the model proposed in \cite{he2022elic}. The kernel size in down/up-sampling layers is changed from 5 to 4 for better performance as suggested in \cite{zhang2023practical}.
    \item \textbf{Shallow-2layer} (ICCV2023) denote models proposed in \cite{yang2023shallow}. We adopt the model with simpler analysis transform and non-residual connection synthesis transform for better performance as suggested in their open-source version\footnote{https://github.com/mandt-lab/shallow-ntc}.
    \item \textbf{TCM} (CVPR2023) denote the model proposed in \cite{liu2023learned}. The setting is $N$=192.
    \item \textbf{FM-intra} (CVPR2024) denotes the intra coding model in \cite{li2024fm}, which is a neural image compression model. The U-Net post-processing is removed to stabilize training.
    \item \textbf{Cheng-ckbd} (CVPR2021) \cite{he2021checkerboard} denotes the Cheng2020 model \cite{cheng2020learned} with Gaussian mixture model and checkerboard context model \cite{he2021checkerboard}. 
\end{itemize}
In FastNIC, Shallow-2layer, ELIC, TCM, and FM-intra models, \textbf{Gaussian Model (GM)} is adopted as the default probabilistic model, and zero-center quantization is applied (\textit{i.e.} coding $\lfloor y-\mu \rceil$). These models are used to verify the effectiveness of the switchable priors method in supporting more complex probabilistic models. We use \textbf{Generalized Gaussian Model (GGM)} \cite{zhang2024ggm} and \textbf{Gaussian mixture model (GMM)} as more advanced probabilistic models. We designed zero-center quantization for GMM for better performance, which is included in Sec. \uppercase\expandafter{\romannumeral3} in the supplementary material.

To highlight the necessity of the multi-dimensional prior set, we evaluate with the Cheng-ckbd model, which employs nonzero-center quantization (\textit{i.e.}, coding $\lfloor y \rceil$). In this scenario, the different mean parameters in GMM should also be considered in the switchable priors.

\begin{table*}[!t]
\centering
\caption{Performance evaluation on Kodak dataset of \textbf{FastNIC} model compared to BPG\tnote{1}.}
\begin{threeparttable}
\begin{tabular}{ ccccc|ccccccc} 
\hline
\multirow{2}{*}{Model}&\multirow{2}{*}{Method}& \multirow{2}{*}{\Centerstack[c]{Probabilistic\\model}}&\multicolumn{2}{c|}{Number of CDF} &\multirow{2}{*}{\Centerstack[c]{BD-\\Rate (\%)}} & \multicolumn{2}{c}{Time (ms)}& \multirow{2}{*}{\Centerstack[c]{CDF table\\size (MB)}}&\multirow{2}{*}{\Centerstack[c]{Net\\params (M)}} &  \multicolumn{2}{c}{KMACs/pixel}\\
&&&$\hat{y}$&$\hat{z}$&&Enc.\tnote{2}&Dec.\tnote{2}&&&Enc.\tnote{2}&Dec.\tnote{2}\\
\hline
\multirow{12}{*}{FastNIC}&\multirow{3}{*}{\Centerstack[c]{Dynamically\\computed CDF}}&GM&393216&192&-2.22&70.6&67.9&63.25&3.6&11.2&9.3\\
&&GMM&393216&192&-3.07&80.5&74.3&68.23&4.1&13.1&11.3\\
&&GGM&393216&192&-3.52&77.1&71.4&67.67&3.7&11.3&9.5\\
\cline{2-12}
&\multirow{3}{*}{\Centerstack[c]{LUTs-based}}&GM&160&192&-2.21&37.7&35.7&0.17&3.6&11.2&9.3\\
&&GMM\tnote{3}&-&-&-&-&-&-&-&-&-\\
&&GGM&12800&192&-3.52&43.5&44.6&6.34&3.7&11.3&9.5\\
\cline{2-12}
&\multirow{3}{*}{\Centerstack[c]{Switch\\w/o skip\tnote{4}}}&GM&40&0&-2.23&34.6&32.2&0.01&3.6&11.2&9.3\\
&&GMM&40&0&-3.09&34.3&32.6&0.01&3.6&11.2&9.3\\
&&GGM&40&0&-3.51&33.9&32.5&0.01&3.6&11.2&9.3\\
\cline{2-12}
&\multirow{3}{*}{Switch\tnote{4}}&GM&40&0&-3.05&17.8&14.3&0.01&3.7&11.3&9.5\\
&&GMM&40&0&-3.29&18.1&14.8&0.01&3.7&11.3&9.5\\
&&GGM&40&0&\textbf{-4.10}&17.2&15.0&0.01&3.7&11.3&9.5\\
\hline
\end{tabular}
\begin{tablenotes}
\footnotesize
\item[1] The anchor for calculating BD-rate is the performance of BPG version 0.9.8. All models are trained for MSE and evaluated with PSNR.
\item[2] ``Encoding" is abbreviated as ``Enc." ``Decoding" is abbreviated as ``Dec."
\item[3] For GMM with 9 parameters, applying the LUTs-based method would incur excessive storage costs. Therefore, we did not implement the LUTs-based method for GMM. The detailed reasons are explained in Sec. \uppercase\expandafter{\romannumeral2} in the supplementary material.
\item[4] ``Switch" denotes our proposed switchable priors method. ``Switch w/o skip" denotes the switchable priors method without skip mode. Both Switch and Switch w/o skip reuse the learned prior set for coding $\hat{z}$.
\end{tablenotes} 
\end{threeparttable}
\label{tab:fastnic_main}
\end{table*}

\begin{figure}[!t]
  \centering
  \subfloat[Trained for MSE]
    {  
        \includegraphics[width=0.46\linewidth]{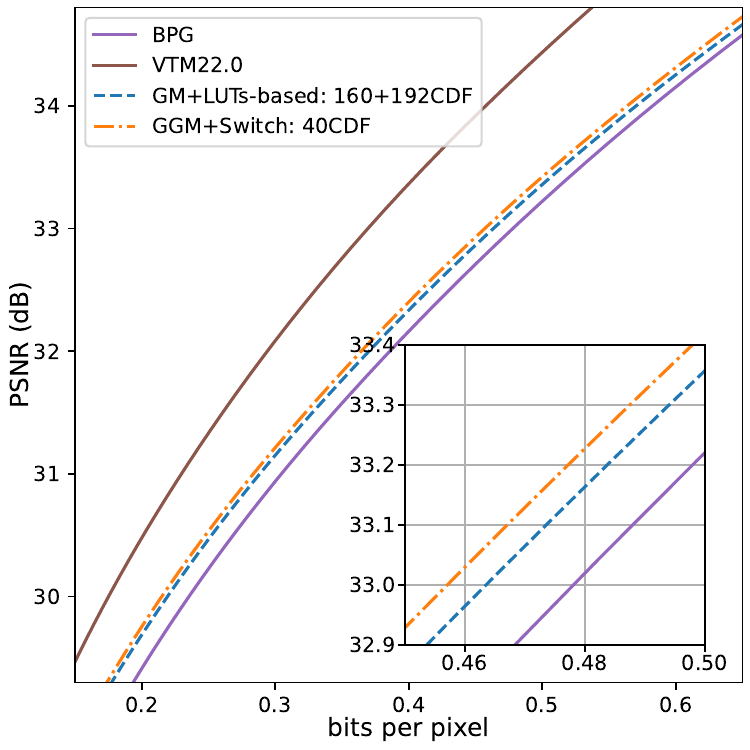}
        \label{fig:fastnic_mse}
    }
    % \hspace{-0.2cm}
    \subfloat[Trained for MS-SSIM]
    {
        \includegraphics[width=0.46\linewidth]{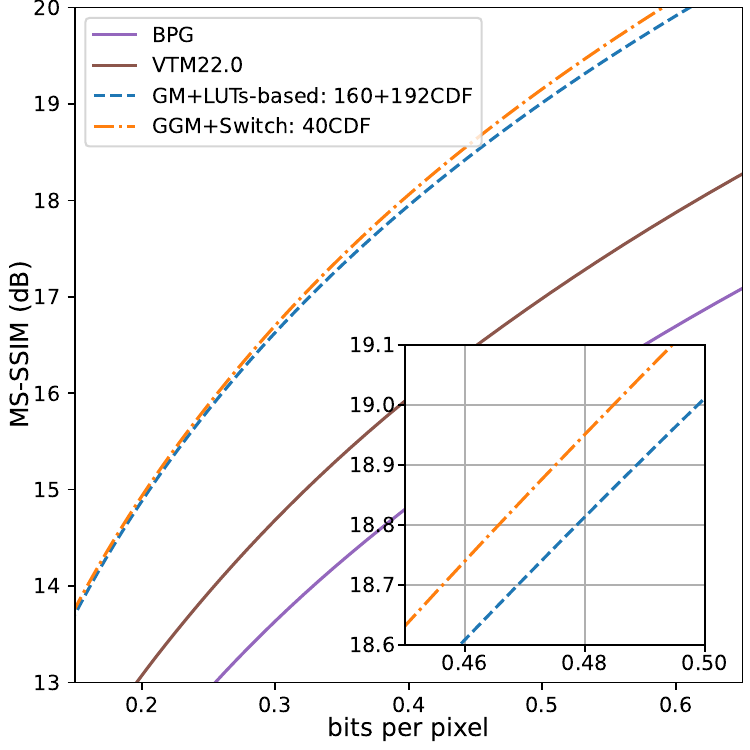}
        \label{fig:fastnic_msssim}
    } 
    %\vspace{-0.2em}
    \caption{Performance of our proposed \textbf{FastNIC+Switch}. (a) shows the performance of FastNIC+Switch trained for MSE. (b) shows the performance of FastNIC+Switch trained for MS-SSIM. ``Switch'' represents our proposed switchable priors method.
    }
    \label{fig:fastnic_rd}
\end{figure}

\subsubsection{Training} For models trained for mean squared error (MSE), we trained multiple models with different values of $\lambda \in \{0.0018, 0.0054, 0.0162, 0.0483\}$. For models trained for multi-scale structural similarity (MS-SSIM), we trained multiple models with different values of $\lambda \in \{2.4,7.2,21.6,64.8\}$. The training dataset is the Flicker2W dataset \cite{liu2020unified}. In each training iteration, images were randomly cropped into 256×256 patches.
The training process utilized the Adam optimizer \cite{kingma2014adam} for 500 epochs, with a batch size of 8 and an initial learning rate of $10^{-4}$. After 400 epochs, the learning rate was reduced to $10^{-5}$, and after 50 more epochs, it was further reduced to $10^{-6}$. All experiments were conducted with the same random seed to ensure consistency.

For training models with switchable priors, we initialized with the anchor model after 400 epochs and then applied the Adam optimizer for an additional 100 epochs for a fair comparison. The initial learning rate was $10^{-4}$. After 50 epochs, the learning rate was reduced to $10^{-5}$, and after 30 epochs, it was further reduced to $10^{-6}$. The annealing schedule for the soft assignment was defined as $\tau = \tau_0 \exp(-0.01 \times \text{epoch})$, where $\tau_0 = 0.05 \times M$, and $M$ is the number of distributions in the prior set. The parameters of the distributions were initialized with an entropy-increasing strategy to facilitate convergence. For example, for the Gaussian model, each distribution is parameterized by $\sigma$, and the $\sigma$ increases as the index of the prior increases. We initialized with the switchable priors model to incorporate the skip mode and optimized only the entropy model for another 100 epochs. The annealing schedule for the skip mode was set as $t = 0.4 \exp(-0.01 \times \text{epoch})$, and the coefficient in the Gumbel-Softmax was fixed as 0.5.

\subsubsection{Implementation}
For the LUTs-based method, we follow the implementation in \cite{balle2019integer,begaint2020compressai,zhang2024ggm}. For GM, the scale values are log-scaled sampled from $[0.11, 60]$. For GGM, the shape values are uniformly sampled from $[0.05,3]$ and the scale values are log-scaled sampled from $[0.01, 60]$. The detailed sampling strategy for LUTs-based entropy coding is included in Sec. \uppercase\expandafter{\romannumeral2} in the supplementary material.

\subsubsection{Evaluation}
We evaluated different methods on four commonly used test datasets. The Kodak dataset \footnote{Available online at \url{http://r0k.us/graphics/kodak/}.} contains 24 images with either 512×768 or 768×512 pixels. The Tecnick dataset \cite{asuni2014testimages} contains 100 images with 1200×1200 pixels. The CLIC 2020 professional validation dataset\footnote{http://compression.cc}, referred to as CLIC, includes 41 high-quality images. The USTC-TD dataset \cite{li2024ustc} contains 40 images with 4K resolution.

We used bits-per-pixel (bpp) and Peak Signal-to-Noise Ratio (PSNR) or MS-SSIM in dB measured in the RGB space to quantify rate and distortion.
We also used the BD-Rate metric \cite{bjontegaard2001calculation} to calculate the rate saving. The BD-Rate for an entire dataset is calculated by averaging the BD-Rate of each image.
We tested the performance of BPG-0.9.8\footnote{https://bellard.org/bpg} and VTM-22.0\footnote{https://vcgit.hhi.fraunhofer.de/jvet/VVCSoftware\_VTM} with input format as YUV444. 
The coding time was tested on a single-core Intel(R) Xeon(R) Gold 6248R CPU @ 3.00GHz and one NVIDIA GeForce RTX 3090 GPU. The coding time is averaged across four models targeting different bitrates.
The complexity measures, such as KMACs/pixel and parameter counts are evaluated using DeepSpeed\footnote{https://github.com/microsoft/DeepSpeed}.

\begin{figure*}[!t]
\centering
\subfloat[Shallow-2layer (ICCV2023)]
{  
    \includegraphics[width=0.24\linewidth]{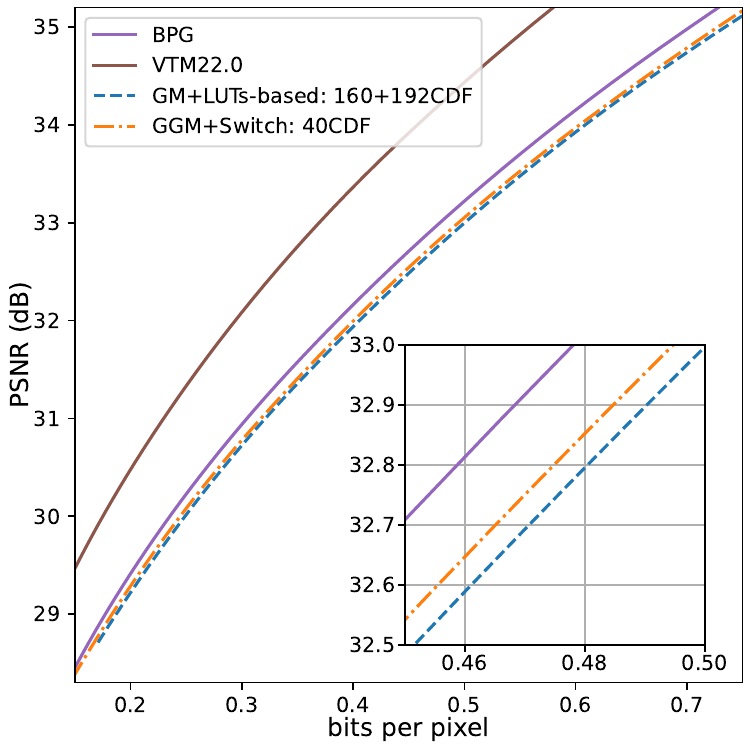}
    \label{fig:shallow_rd}
} 
\subfloat[ELIC (CVPR2022)]
{  
    \includegraphics[width=0.24\linewidth]{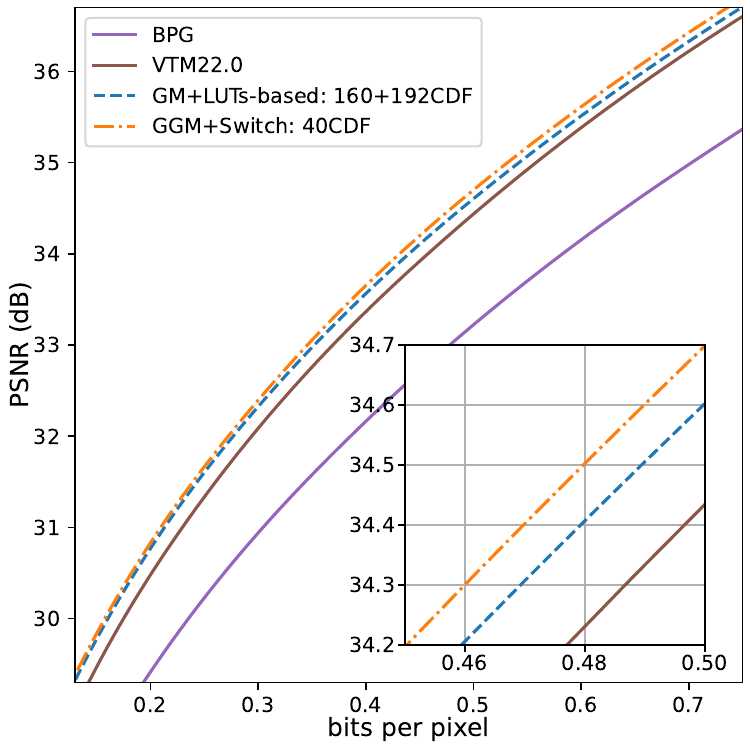}
    \label{fig:elic_rd}
} 
\subfloat[TCM (CVPR 2023)]
{  
    \includegraphics[width=0.24\linewidth]{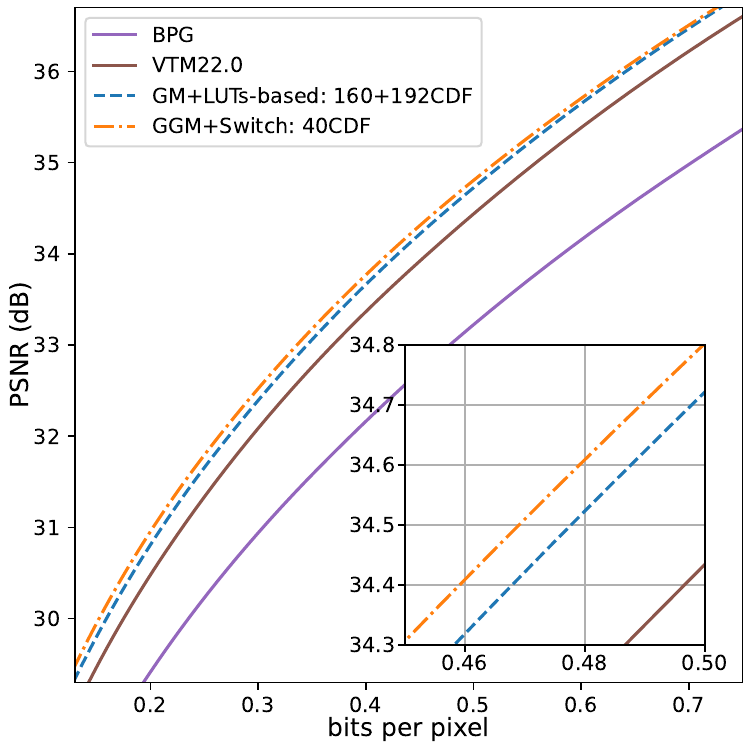}
    \label{fig:tcm_rd}
} 
\subfloat[FM-intra (CVPR2024)]
{  
    \includegraphics[width=0.24\linewidth]{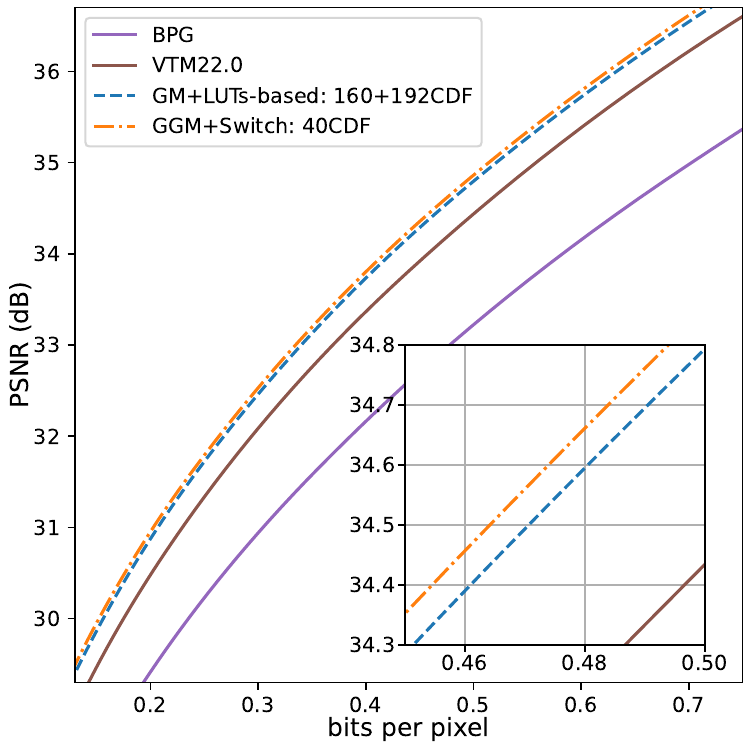}
    \label{fig:fm_rd}
} 
\caption{
Rate-distortion performance of different methods on the Kodak dataset. All models are trained for MSE and evaluated with PSNR. ``Switch'' represents our proposed switchable priors method.
}
\label{fig:rd}
\end{figure*}

\begin{table*}[!t]
\centering
\caption{Performance evaluation on Kodak dataset of different methods.}
\begin{threeparttable}
\begin{tabular}{ ccccc|ccccccc} 
\hline
\multirow{2}{*}{Model}&\multirow{2}{*}{Method}& \multirow{2}{*}{\Centerstack[c]{Probabilistic\\model}}&\multicolumn{2}{c|}{Num. CDF\tnote{2}} &\multirow{2}{*}{\Centerstack[c]{BD-\\Rate (\%)}} & \multicolumn{2}{c}{Time (ms)}& \multirow{2}{*}{\Centerstack[c]{CDF table\\size (MB)}}&\multirow{2}{*}{\Centerstack[c]{Net\\params (M)}} &  \multicolumn{2}{c}{KMACs/pixel}\\
&&&$\hat{y}$&$\hat{z}$&&Enc.\tnote{2}&Dec.\tnote{2}&&&Enc.&Dec.\\
\hline
\multirow{4}{*}{\Centerstack[c]{Shallow-\\2layer\cite{yang2023shallow}\\(ICCV2023)}}&\multirow{2}{*}{LUTs-based}&GM&160 &192 &0.00 &42.8 &39.5 &0.17 &19.0 &165.1 &19.0 \\
&&GGM&12800 &192 &-1.95 &43.2 &43.8 &6.34 &19.1 &165.6 &19.6 \\
\cline{2-12}
&Switch w/o skip\tnote{3}&GGM&40 &0 &-1.91 &38.7 &35.8 &0.01 &19.0 &165.1 &19.0 \\
&Switch\tnote{3}&GGM&40 &0 &\textbf{-2.60}&14.9 &11.2 &0.01 &19.1 &165.6 &19.6 \\
\hline
\multirow{4}{*}{\Centerstack[c]{Our\\FastNIC}}&\multirow{2}{*}{LUTs-based}&GM&160 &192 &0.00 &37.7 &35.7 &0.17 &3.6 &11.2 &9.3 \\
&&GGM&12800 &192 &-1.34 &43.5 &44.6 &6.34 &3.7 &11.3 &9.5 \\
\cline{2-12}
&Switch w/o skip&GGM&40 &0 &-1.33 &33.9 &32.5 &0.01 &3.6 &11.2 &9.3 \\
&Switch&GGM&40 &0 &\textbf{-1.90}&17.2 &15.0 &0.01 &3.7 &11.3 &9.5 \\
\hline
\multirow{4}{*}{\Centerstack[c]{ELIC\cite{he2022elic}\\(CVPR2022)}}&\multirow{2}{*}{LUTs-based}&GM&160 &192 &0.00 &78.5 &86.7 &0.17 &33.0 &306.1 &303.0 \\
&&GGM&12800 &192 &-2.09 &85.2 &90.4 &6.34 &33.1 &306.6 &303.7 \\
\cline{2-12}
&Switch w/o skip&GGM&40 &0 &-2.11 &71.9 &78.9 &0.01 &33.0 &306.1 &303.0 \\
&Switch&GGM&40 &0 &\textbf{-2.34}&50.6 &55.2 &0.01 &33.1 &306.6 &303.7 \\
\hline
\multirow{4}{*}{\Centerstack[c]{TCM\cite{liu2023learned}\\(CVPR2023)}}&\multirow{2}{*}{LUTs-based}&GM&160 &192 &0.00 &82.7 &89.0 &0.17 &65.2 &529.4 &733.7 \\
&&GGM&12800 &192 &-2.68 &92.6 &101.3&6.34 &65.3 &529.7 &733.9 \\
\cline{2-12}
&Switch w/o skip&GGM&40 &0 &-2.61 &72.1 &82.6 &0.01 &65.2 &529.4 &733.7 \\
&Switch&GGM&40 &0 &\textbf{-2.69}&49.3 &54.4 &0.01 &65.3 &529.7 &733.9 \\
\hline
\multirow{4}{*}{\Centerstack[c]{FM-intra\cite{li2024fm}\\(CVPR2024)}}&\multirow{2}{*}{LUTs-based}&GM&160 &192 &0.00 &64.6 &67.6 &0.17 &41.0 &387.4 &385.1 \\
&&GGM&12800 &192 &-1.71 &74.4 &80.7 &6.34 &41.2 &390.6 &388.6 \\
\cline{2-12}
&Switch w/o skip&GGM&40 &0 &-1.75 &55.0 &61.8 &0.01 &41.0 &387.4 &385.1 \\
&Switch&GGM&40 &0 &\textbf{-1.87}&33.4 &34.9 &0.01 &41.2 &390.6 &388.6 \\
\hline
\end{tabular}
\begin{tablenotes}
\footnotesize
\item[1] The anchor for calculating BD-rate for each model is the performance of using GM and LUTs-based method. All models are trained for MSE and evaluated with PSNR.
\item[2] ``Number of CDF" is abbreviated as ``Num. CDF." ``Encoding" is abbreviated as ``Enc." ``Decoding" is abbreviated as ``Dec."
\end{tablenotes} 
\end{threeparttable}
\label{tab:main_bd}
\end{table*}

\begin{figure*}[!t]
\centering
\subfloat[]
{  
    \includegraphics[width=0.253\linewidth]{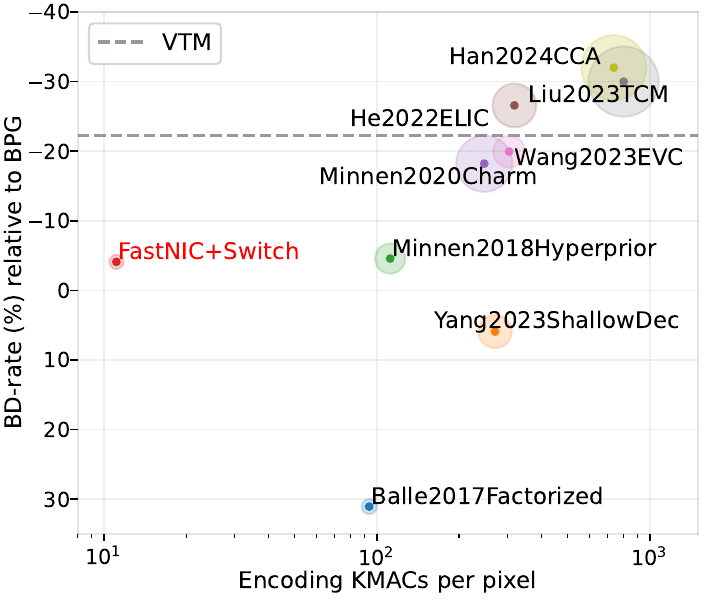}
    \label{fig:mac_enc}
}\hspace{-0.35cm}
\subfloat[]
{  
    \includegraphics[width=0.238\linewidth]{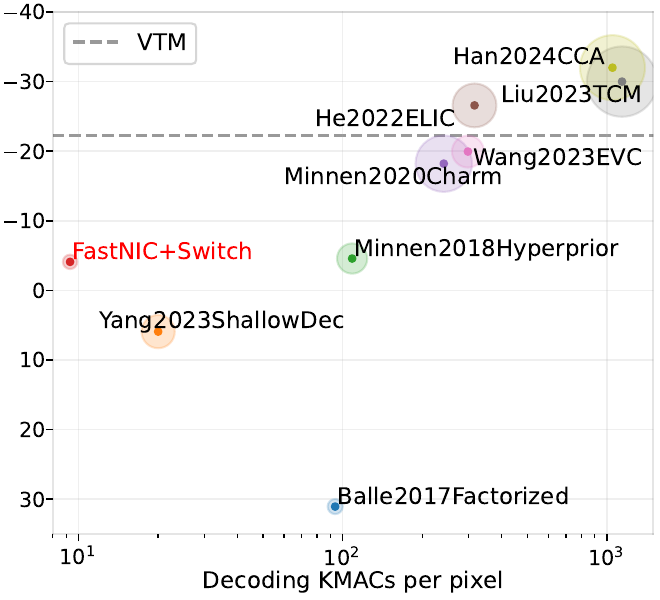}
    \label{fig:mac_dec}
} \hspace{-0.35cm}
\subfloat[]
{  
    \includegraphics[width=0.238\linewidth]{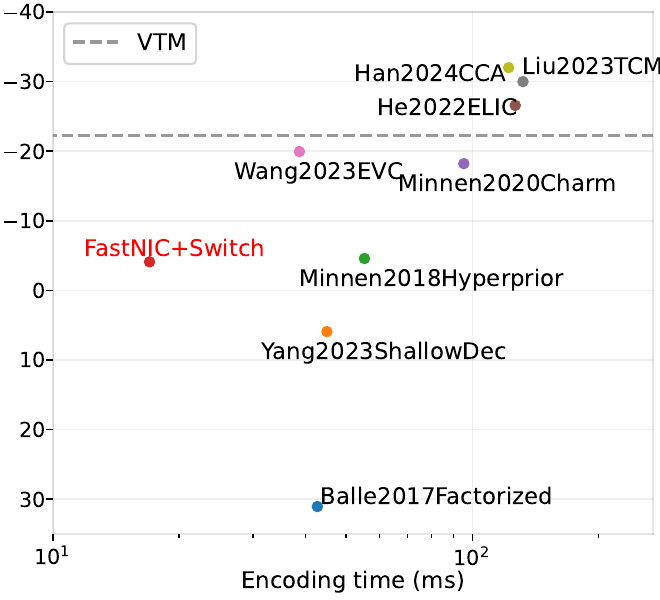}
    \label{fig:time_enc}
} \hspace{-0.35cm}
\subfloat[]
{  
    \includegraphics[width=0.238\linewidth]{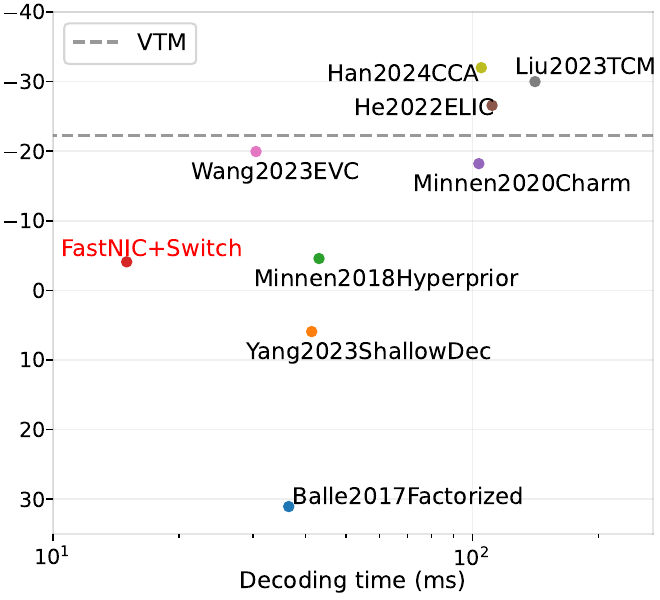}
    \label{fig:time_dec}
} 
\caption{
Performance comparison of the proposed FastNIC+Switch model with other NIC models on the Kodak dataset. All models are trained for MSE and evaluated with PSNR. The $x$-axis is log-scaled. In (a) and (b), the circle radius corresponds to the parameter count of each method. 
}
\label{fig:vs}
\end{figure*}

\subsection{Compression Performance and Complexity}
We abbreviate the switchable priors method incorporating the skip mode as \textbf{Switch} and the basic switchable priors method without the skip mode as \textbf{Switch w/o skip}. Table \ref{tab:fastnic_main} and \ref{tab:main_bd} show the compression performance and complexity. 

We conducted a thorough comparison between GM, GMM, and GGM.  It is worth noting that our focus is on demonstrating that our switchable priors method can support more complex probabilistic models, including GMM and GGM, without increasing complexity. The advantage of these complex probabilistic models is not a contribution of our study. As shown in Table \ref{tab:fastnic_main}, the experimental results indicate that regardless of the entropy coding methods used, GGM consistently outperforms both GM and GMM. Therefore, we primarily use GGM to validate the effectiveness of our proposed switchable priors method in supporting more complex probabilistic models.

As shown in Table \ref{tab:fastnic_main}, applying the more complex GGM results in better compression performance compared to the Gaussian model. However, both dynamically calculating CDF tables and LUTs-based implementation introduce higher complexity when applying GGM. Specifically, first, the network complexity increases due to more parameters that need to be predicted. Second, when dynamically calculating CDF tables, the coding time increases due to the higher computational cost of constructing the CDF tables. Third, with the LUTs-based method, GGM requires more CDF tables for entropy coding, leading to longer memory access time and higher computational costs for building indexes to the corresponding CDF tables, which in turn increases coding time.

In contrast, our Switch w/o skip supports GGM with significantly fewer CDF tables for entropy coding compared to the LUTs-based method. Specifically, Switch w/o skip requires only 40 CDF tables to achieve compression performance comparable to the LUTs-based method. This reduction in the number of CDF tables leads to lower storage costs and shorter coding time. Additionally, since the output dimension of the entropy parameters module is smaller, the neural network complexity is slightly reduced.

After incorporating the skip mode, our Switch method further reduces coding time and slightly improves compression performance. However, since the entropy parameters module needs to output a binary mask for skipping, the parameter count and computational complexity of the neural network increase slightly compared to Switch w/o skip. As shown in Table \ref{tab:main_bd}, on ELIC, TCM, and FM-intra models, which have more accurate distribution modeling capabilities brought by context models, the compression performance gain from incorporating the skip mode is smaller than that observed in our FastNIC model.

The rate-distortion curves on the Kodak dataset, shown in Fig. \ref{fig:fastnic_rd} and Fig. \ref{fig:rd}, demonstrate consistent improvements across various bitrate ranges. The experimental results shown in Table \ref{tab:main_bd} demonstrate that our proposed switchable priors method could improve compression performance and reduce computational complexity and coding time on a variety of representative neural image compression models.

As shown in Table \ref{tab:fastnic_main}, when trained for MSE, our FastNIC+Switch achieves a 4.10\% rate saving compared to BPG on the Kodak dataset. As shown in Fig. \ref{fig:vs}, among the NIC methods that outperform BPG, FastNIC achieves the lowest computational complexity and the fastest coding speed, demonstrating its potential for practical deployment. Methods that rely on iterative per-sample optimization are not included in the comparison due to their excessive encoding times. Furthermore, as shown in Fig. \ref{fig:fastnic_msssim}, when trained for MS-SSIM, FastNIC exhibits significantly better performance than VTM, highlighting its practical value when perceptual quality is concerned.

\begin{table*}[!t]
\centering
\caption{Compression performance and coding time of \textbf{FastNIC} model with different methods.}
%\vspace{-0.5em}
\begin{threeparttable}
\begin{tabular}{ c|c|cc|ccc|ccc|ccc|ccc} 
\hline
\multirow{3}{*}{Method}& \multirow{3}{*}{\Centerstack[c]{Prob.\\model\tnote{2}}}  & \multicolumn{2}{c|}{\multirow{2}{*}{\Centerstack[c]{Number\\ of CDF}}} & \multicolumn{3}{c|}{Kodak 512×768} & \multicolumn{3}{c|}{Tecnick 1200×1200} & \multicolumn{3}{c|}{CLIC 1200×1800} & \multicolumn{3}{c}{USTC-TD 2160×4096}\\
&&&&\multirow{2}{*}{\Centerstack[c]{BD-\\rate (\%)}}&\multirow{2}{*}{\Centerstack[c]{$t_e$\tnote{3}\\(ms)}}&\multirow{2}{*}{\Centerstack[c]{$t_d$\tnote{3}\\(ms)}}&\multirow{2}{*}{\Centerstack[c]{BD-\\rate (\%)}}&\multirow{2}{*}{\Centerstack[c]{$t_e$\\(ms)}}&\multirow{2}{*}{\Centerstack[c]{$t_d$\\(ms)}}&\multirow{2}{*}{\Centerstack[c]{BD-\\rate (\%)}}&\multirow{2}{*}{\Centerstack[c]{$t_e$\\(ms)}}&\multirow{2}{*}{\Centerstack[c]{$t_d$\\(ms)}}&\multirow{2}{*}{\Centerstack[c]{BD-\\rate (\%)}}&\multirow{2}{*}{\Centerstack[c]{$t_e$\\(ms)}}&\multirow{2}{*}{\Centerstack[c]{$t_d$\\(ms)}}\\
&&$\hat{y}$&$\hat{z}$&&&&&&&&&&&&\\
\hline
\multirow{11}{*}{\Centerstack[c]{LUTs\\-based}}&\multirow{6}{*}{\Centerstack[c]{GM}}&5&192&62.64&34.2&32.6&75.45&108.6&111.7&61.48&170.1&169.3&82.35&626.5&608.2\\
&&10&192&8.99&34.6&32.9&15.36&108.3&111.6&5.29&169.8&169.6&15.50&620.8&610.0\\
&&20&192&-0.04&34.7&32.8&6.47&108.8&111.3&-3.19&171.0&170.0&6.63&626.8&604.4\\
&&40&192&-1.71&35.2&33.2&4.67&109.6&112.2&-4.92&171.6&169.5&4.90&637.7&610.9\\
&&80&192&-2.12&35.5&33.6&4.29&111.7&113.9&-5.30&174.4&172.7&4.53&639.1&624.0\\
&&160&192&-2.21&37.7&35.7&4.22&115.4&117.1&-5.38&180.2&178.7&4.46&658.8&636.2\\
\cline{2-16}
&\multirow{5}{*}{\Centerstack[c]{GGM}}&50&192&19.30&33.3&34.9&34.94&103.1&114.9&24.85&169.4&174.9&48.05&624.6&627.1\\
&&200&192&-0.97&35.0&35.2&4.76&109.7&114.5&-4.92&168.5&173.2&4.45&622.4&621.8\\
&&800&192&-3.07&35.0&37.4&2.64&110.5&115.3&-6.67&170.8&176.0&2.61&623.8&624.6\\
&&3200&192&-3.46&37.4&38.6&2.26&115.7&120.8&-6.99&174.3&183.4&2.26&636.5&643.7\\
&&12800&192&-3.52&43.5&44.6&2.20&119.4&126.5&-7.04&183.6&192.6&2.22&640.2&657.5\\
\hline
\multirow{4}{*}{\Centerstack[c]{Switch\\w/o skip}}&\multirow{4}{*}{\Centerstack[c]{GGM}}&5&0&0.09&34.2&32.2&6.75&105.1&106.2&-3.03&165.6&164.4&6.86&612.7&591.8\\
&&10&0&-2.66&33.7&32.1&3.31&104.6&107.0&-6.01&167.1&166.0&3.44&612.9&598.0\\
&&20&0&-3.34&33.4&32.0&2.38&104.6&106.9&-6.79&168.1&168.3&2.53&612.7&595.5\\
&&40&0&-3.51&33.9&32.5&2.13&105.1&107.4&-7.03&165.3&164.7&2.26&616.3&596.7\\
\hline
\multirow{4}{*}{\Centerstack[c]{Switch}}&\multirow{4}{*}{\Centerstack[c]{GGM}}&5&0&-2.15&15.9&13.6&5.05&29.7&30.1&-4.49&42.1&41.6&6.17&102.1&99.4\\
&&10&0&-3.58&16.6&14.5&2.82&30.5&30.9&-6.53&42.6&42.3&3.49&104.8&101.9\\
&&20&0&-4.03&17.1&15.0&1.95&30.5&30.9&-7.34&43.0&42.8&2.40&105.0&102.3\\
&&40&0&\textbf{-4.10}&17.2&15.0&\textbf{1.67}&30.5&31.0&\textbf{-7.61}&43.5&43.5&\textbf{2.00}&105.2&102.8\\
\hline
\end{tabular}
\begin{tablenotes}
\footnotesize
\item[1] BD-rate$\downarrow$ is compared to the performance of BPG. All models are trained for MSE and evaluated with PSNR. 
\item[2] ``Probabilistic model" is abbreviated as ``Prob. model." 
\item[3] $t_e$ and $t_d$ represent the encoding and decoding time, respectively. The time presented in this table includes entropy coding time.
\end{tablenotes} 
\end{threeparttable}
\label{tab:fastnic_num_cdf}
\end{table*}

\begin{table*}[!t]
\centering
\caption{Encoding time of different methods based on \textbf{FastNIC}.}
\begin{threeparttable}
\begin{tabular}{ cccc|cccccc|cc} 
\hline
\multirow{3}{*}{\Centerstack[c]{Method}}&\multirow{3}{*}{\Centerstack[c]{Probabilistic\\model}}&\multicolumn{2}{c|}{\multirow{2}{*}{\Centerstack[c]{Number\\of CDF}}}& \multicolumn{8}{c}{\Centerstack[c]{Time (ms)}} \\
\cline{5-12}
&&&&\multirow{2}{*}{\Centerstack[c]{$g_a$\tnote{1}}}&\multirow{2}{*}{\Centerstack[c]{$h_a$\tnote{1}}}&\multirow{2}{*}{\Centerstack[c]{$h_s$\tnote{1}}}&\multirow{2}{*}{\Centerstack[c]{$\text{EE}_z$\tnote{2}}}&\multirow{2}{*}{\Centerstack[c]{$\text{EE}_y$\tnote{2}}}&\multirow{2}{*}{\Centerstack[c]{Total\\w/o build index}}&\multirow{2}{*}{\Centerstack[c]{build\\index}}&\multirow{2}{*}{\Centerstack[c]{Total\\w/build index}} \\
&&\Centerstack[c]{$\hat{y}$}&\Centerstack[c]{$\hat{z}$}&&&&&&&&\\
\hline
\multirow{2}{*}{LUTs-based}&GM&160&192&2.2&1.3&2.6&1.6&26.8&34.5&3.0&37.5\\
&GGM&12800&192&2.2&1.3&2.6&1.6&36.7&44.4&3.6&48.0\\
\hline
\multirow{1}{*}{Switch w/o skip}&GGM&40&0&2.1&1.3&2.5&1.4&25.5&32.8&0.1&32.9\\
\multirow{1}{*}{Switch}&GGM&40&0&2.2&1.4&2.7&1.1&10.3&17.7&0.2&17.9\\
\hline
\end{tabular}
\begin{tablenotes}
\footnotesize
\item[1] ``$g_a$'', ``$h_a$'', and ``$h_s$'' denote the network inference time of the analysis transform, hyper analysis transform, and hyper synthesis transform, respectively.
\item[2] ``$\text{EE}_z$'' and ``$\text{EE}_y$'' denote the entropy encoding time for hyper latents and main latents, respectively.
\item[3] The results are evaluated on the Kodak dataset.
\end{tablenotes} 
\end{threeparttable}
\label{tab:enc_time}
\end{table*}

\begin{table*}[!t]
\centering
\caption{Decoding time of different methods based on \textbf{FastNIC}.}
\begin{threeparttable}
\begin{tabular}{ cccc|ccccc|cc} 
\hline
\multirow{3}{*}{\Centerstack[c]{Method}}&\multirow{3}{*}{\Centerstack[c]{Probabilistic\\model}}&\multicolumn{2}{c|}{\multirow{2}{*}{\Centerstack[c]{Number\\of CDF}}}& \multicolumn{7}{c}{\Centerstack[c]{Time (ms)}} \\
\cline{5-11}
&&&&\multirow{2}{*}{\Centerstack[c]{$g_s$\tnote{1}}}&\multirow{2}{*}{\Centerstack[c]{$h_s$\tnote{1}}}&\multirow{2}{*}{\Centerstack[c]{$\text{ED}_z$\tnote{2}}}&\multirow{2}{*}{\Centerstack[c]{$\text{ED}_y$\tnote{2}}}&\multirow{2}{*}{\Centerstack[c]{Total\\w/o build index}}&\multirow{2}{*}{\Centerstack[c]{build\\index}}&\multirow{2}{*}{\Centerstack[c]{Total\\w/ build index}} \\
&&\Centerstack[c]{$\hat{y}$}&\Centerstack[c]{$\hat{z}$}&&&&&&&\\
\hline
\multirow{2}{*}{LUTs-based}&GM&160&192&2.1 &2.7 &1.7 &26.6 &33.1 &3.2 &36.3 \\
&GGM&12800&192&2.2 &2.6 &1.8 &35.7 &42.3 &3.7 &46.0 \\
\hline
\multirow{1}{*}{Switch w/o skip}&GGM&40&0&2.1 &2.5 &1.6 &26.5 &32.7 &0.1 &32.8 \\
\multirow{1}{*}{Switch}&GGM&40&0&2.2 &2.6 &1.2 &8.9 &14.9 &0.2 &15.1 \\
\hline
\end{tabular}
\begin{tablenotes}
\footnotesize
\item[1] ``$g_s$'' and ``$h_s$'' denote the network inference time of the synthesis transform  and hyper synthesis transform, respectively.
\item[2] ``$\text{ED}_z$'' and ``$\text{ED}_y$'' denote the entropy decoding time for hyper latents and main latents. 
\item[3] The results are evaluated on the Kodak dataset.
\end{tablenotes} 
\end{threeparttable}
\label{tab:dec_time}
%\vspace{-1em}
\end{table*}

\subsection{Analyses on Switchable Priors Method}
In this section, we analyze the performance of Switch w/o skip. Due to the learning capability of the prior set, Switch w/o skip significantly outperforms the LUTs-based method when the number of CDF tables is limited, as shown in Tables \ref{tab:fastnic_num_cdf} and \ref{tab:fm_num_cdf}. For GGM, Switch w/o skip with only 40 CDF tables achieves performance comparable to the LUTs-based approach with 12800 CDF tables.
Even with as few as 5 CDF tables, the rate increase is less than 4\% compared to using 40 CDF tables. In addition to supporting advanced probabilistic models, Switch w/o skip also reduces entropy coding time. For both LUTs-based methods and our Switch w/o skip, coding time increases as the number of CDF tables grows. However, the number of CDF tables required by the switchable priors method is significantly smaller, resulting in shorter coding time. Furthermore, the time savings are more pronounced in the FM-intra model, which uses a 4-step context model, where each step involves entropy coding.

The reduction in coding time in our method arises from two main factors. First, our approach significantly decreases the number of CDF tables required for entropy coding. As demonstrated in Table \ref{tab:ablation_switch}, dynamically calculating CDF tables for each element requires substantial memory. For instance, in the proposed FastNIC model, dynamically calculating CDF tables for a single 4K image from the USTC-TD dataset requires generating 8 million CDF tables, resulting in a memory consumption of 1.9GB. The LUTs-based method requires less storage for CDF tables, but the storage cost remains notable. By leveraging the switchable priors and sharing the prior set with hyperlatents, our approach significantly reduces memory usage. This reduction in the number of CDF tables helps accelerate entropy coding by reducing memory access time. Second, our method also speeds up the process of generating CDF table indexes for entropy coding. Dynamically calculating CDF tables is computationally expensive, taking 95ms for a single 4K image. The LUTs-based method \cite{begaint2020compressai} involves a searching and discretization process for index construction, which takes 36ms for a 4K image. In contrast, our switchable priors method generates indexes through a simple discretization operator, requiring only 0.2ms for a 4K image. This reduction in index generation time contributes to the overall speed advantage of our method.

\begin{table}[!t]
\centering
\caption{Cost of constructing CDF tables or building indexes on \textbf{FastNIC} model with GGM.}
\begin{threeparttable}
\begin{tabular}{ c|c|cc} 
\hline
\Centerstack[c]{Method} & \Centerstack[c]{Number\\of CDF} & \Centerstack[c]{CDF table\tnote{1}\\size (B)}&\Centerstack[c]{$t$ (ms)}\tnote{1}\\
\hline
\Centerstack[c]{Dynamically computed CDF}&8M&1.9G&95\\
\hline
\multirow{1}{*}{\Centerstack[c]{LUTs-based}}&12.5K&6.34M&36\\
\hline
\multirow{1}{*}{\Centerstack[c]{Switch w/o skip}}&40&12K&0.2\\
\hline
\end{tabular}
\begin{tablenotes}
\footnotesize
\item[1] The memory and time cost are evaluated on USTC-TD. The time cost only contains the process of calculating CDF tables and building indexes on GPU. The time for transmitting CDF tables from GPU to CPU is not included.
\end{tablenotes} 
\end{threeparttable}
\label{tab:ablation_switch}
%\vspace{-1.2em}
\end{table}

The detailed coding time consumption of different components is shown in Table \ref{tab:enc_time} and Table \ref{tab:dec_time}. The network inference time remains comparable across different methods. Compared to the basic LUTs-based approach, our switchable priors method reduces the time required for index generation and entropy coding. This reduction in entropy coding time is particularly evident for GGM, which typically requires many CDF tables for entropy coding. Our approach can support GGM using only 40 CDF tables, resulting in a significant reduction in entropy coding time for both encoding and decoding processes compared to the LUTs-based approach.

\begin{table}[!t]
\centering
\caption{Compression performance and coding time of \textbf{FM-intra} model with different methods.}
%\vspace{-0.4em}
\begin{threeparttable}
\begin{tabular}{ c|c|cc|ccc} 
\hline
\multirow{3}{*}{Method} & \multirow{3}{*}{\Centerstack[c]{Prob.\\model}} & \multicolumn{2}{c|}{\multirow{2}{*}{\Centerstack[c]{Number\\of CDF}}} & \multicolumn{3}{c}{Kodak}\\
&&&&\multirow{2}{*}{\Centerstack[c]{BD-\\rate (\%)}}&\multirow{2}{*}{\Centerstack[c]{$t_e$\\(ms)}}&\multirow{2}{*}{\Centerstack[c]{$t_d$\\(ms)}}\\
&&$\hat{y}$&$\hat{z}$&&&\\
\hline
\multirow{11}{*}{\Centerstack[c]{LUTs\\-based}}&\multirow{6}{*}{\Centerstack[c]{GM}}&5&192&50.09&54.5&62.0\\
&&10&192&2.51&55.6&62.8\\
&&20&192&-6.69&56.8&63.2\\
&&40&192&-8.37&57.5&62.7\\
&&80&192&-8.71&58.9&63.9\\
&&160&192&-8.91&64.6&67.6\\
\cline{2-7}
&\multirow{5}{*}{\Centerstack[c]{GGM}}&50&192&18.61&57.7&65.0\\
&&200&192&-7.26&58.7&65.1\\
&&800&192&-10.14&62.1&68.9\\
&&3200&192&-10.60&67.8&73.0\\
&&12800&192&-10.65&74.4&80.7\\
\hline
\multirow{4}{*}{\Centerstack[c]{Switch\\w/o skip}}&\multirow{4}{*}{\Centerstack[c]{GGM}}&5&0&-7.50&54.6&61.5\\
&&10&0&-9.90&54.8&61.9\\
&&20&0&-10.46&54.2&61.2\\
&&40&0&-10.60&55.0&61.8\\
\hline
\multirow{4}{*}{\Centerstack[c]{Switch}}&\multirow{4}{*}{\Centerstack[c]{GGM}}&5&0&-8.74&31.0&32.1\\
&&10&0&-10.18&31.9&33.2\\
&&20&0&-10.61&32.3&33.7\\
&&40&0&\textbf{-10.71}&33.4&34.9\\
\hline
\end{tabular}
\begin{tablenotes}
\footnotesize
\item[1] BD-rate$\downarrow$ is compared to VTM22.0. All models are trained for MSE and evaluated with PSNR.
\end{tablenotes} 
\end{threeparttable}
\label{tab:fm_num_cdf}
%\vspace{-1.2em}
\end{table}
\subsection{Analyses on Skip Mode}
\begin{figure}[!t]
\centering
\subfloat[]
{  
    \includegraphics[width=0.48\linewidth]{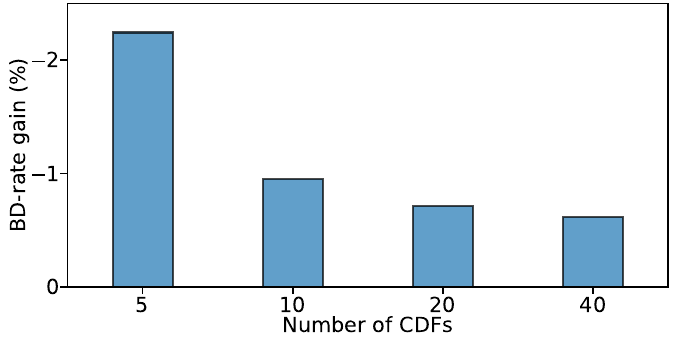}
    \label{fig:bd_wrt_CDF}
}\hspace{-0.3cm}
\subfloat[]
{  
    \includegraphics[width=0.48\linewidth]{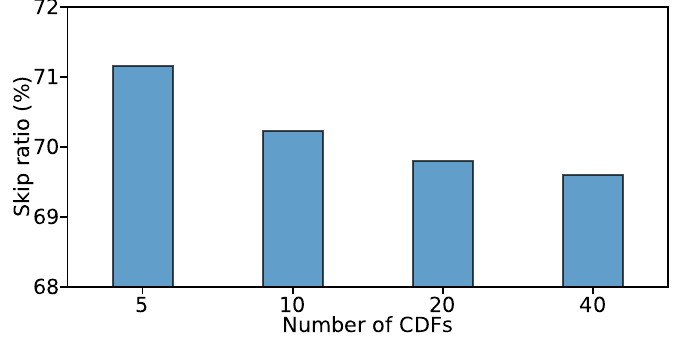}
    \label{fig:skipratio_wrt_CDF}
} 
%\vspace{-0.2em}
\caption{Performance of incorporating skip mode concerning the number of CDF tables. (a) shows the BD-rate gain. (b) shows the skip ratio. The results are collected from the Kodak set on the FastNIC model.
}
\label{fig:skip_wrt_cdf}
%\vspace{-1em}
\end{figure}

In this section, we analyze the performance of the skip mode. Incorporating the skip mode leads to improved compression performance, as shown in Table \ref{tab:fastnic_main} and Table \ref{tab:main_bd}. The primary factor influencing the coding gain of the skip mode is the number of CDF tables used. When fewer CDF tables are employed, the accuracy of distribution modeling decreases. This reduction in accuracy can lead to a higher rate overhead for some latent variables. Consequently, introducing a skipping mechanism when fewer CDF tables are employed can bring larger performance improvements. As shown in Fig. \ref{fig:bd_wrt_CDF}, the coding gain of the skip mode increases as the number of CDF tables is reduced. Additionally, the skip ratio also increases with a reduction in the number of CDF tables, as shown in Fig. \ref{fig:skipratio_wrt_CDF}.

\begin{figure}[!t]
\centering
% %\vspace{-0.8em}
\subfloat[Avg. bpp: 0.13($\lambda=0.0018$)]
{  
    \includegraphics[width=0.48\linewidth]{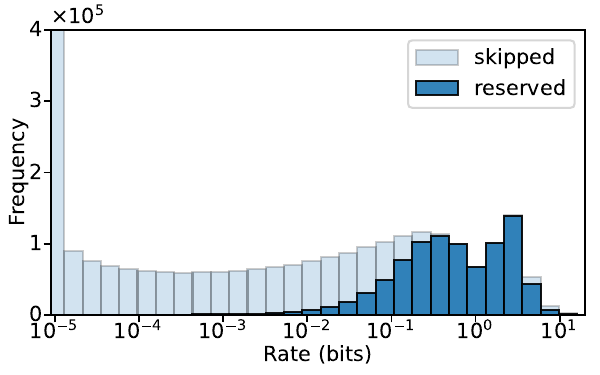}
    \label{fig:fre_low}
}\hspace{-0.3cm}
\subfloat[Avg. bpp: 0.98($\lambda=0.0483$)]
{  
    \includegraphics[width=0.48\linewidth]{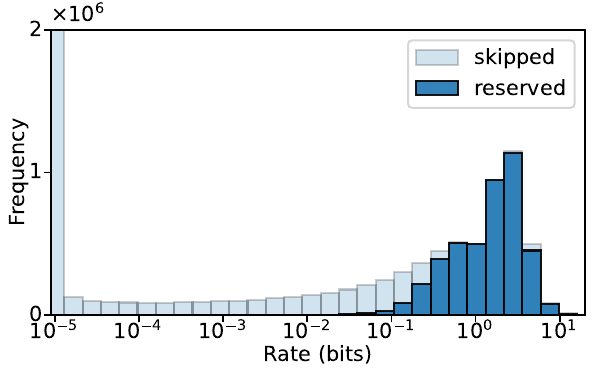}
    \label{fig:fre_high}
} 

%\vspace{-0.2em}
\subfloat[Skip ratio.]
{  
    \includegraphics[width=0.9\linewidth]{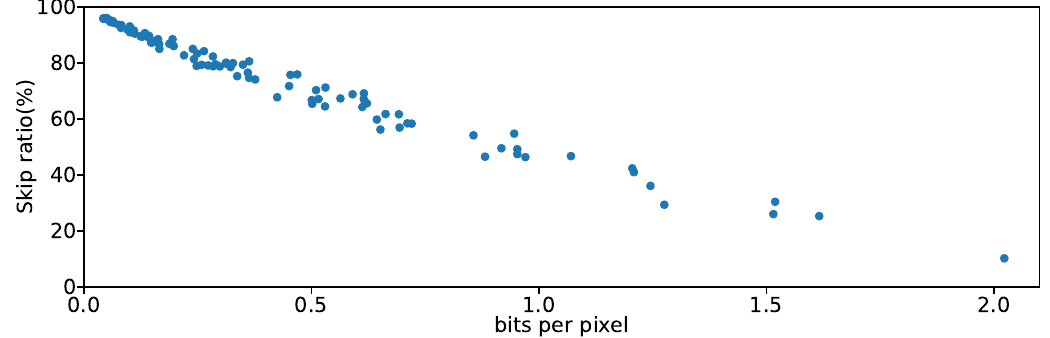}
    \label{fig:skip_bpp}
} 
%\vspace{-0.2em}
\caption{
Skip ratio of latents in FastNIC model. (a) and (b) shows the frequency of the rate of skipped and reserved latents. (c) shows the relationship between the bitrate and skip ratio. The results are collected from the Kodak set on the FastNIC model.
}
\label{fig:skip_ratio}
%\vspace{-0.3em}
\end{figure}

\begin{figure}[!t]
  \centering
  \subfloat[Rate saving at different bitrates.]
    {  
        \includegraphics[width=0.48\linewidth]{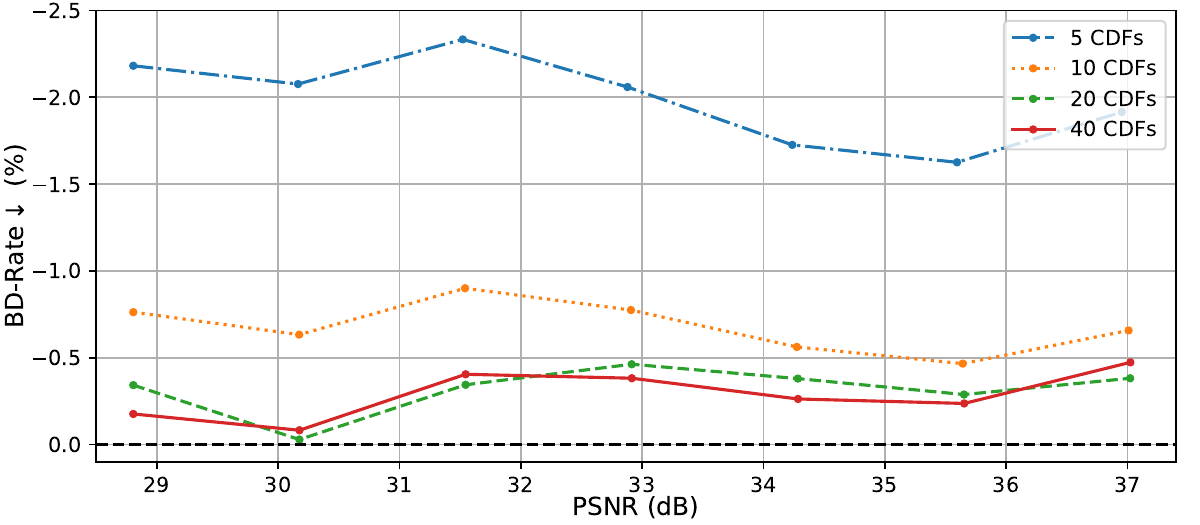}
        \label{fig:rate_saving_rate}
    }
  \subfloat[Rate saving at different skip ratios.]
    {  
        \includegraphics[width=0.48\linewidth]{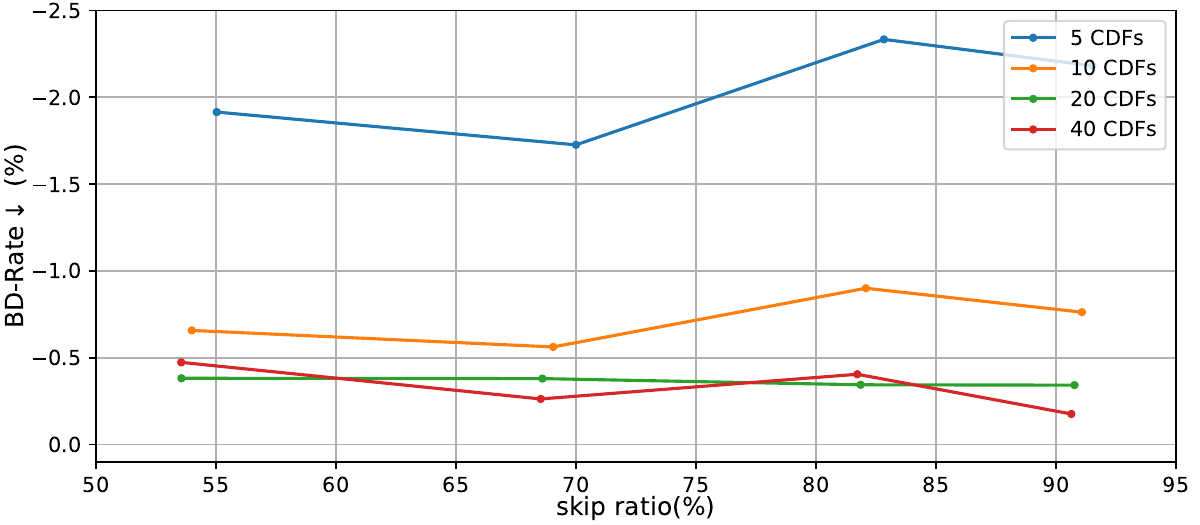}
        \label{fig:rate_saving_ratio}
    } 
    
    \caption{
    Rate saving at different bitrates and skip ratios. The results are collected from the Kodak set on the FastNIC model.
    }
    \label{fig:rate_saving}
\end{figure}

As shown in Fig. \ref{fig:rate_saving}, the performance improvement brought by the skip mode does not significantly vary across different bitrates or different skip ratios. We discuss the reasons in the following. The decision of skip mode is optimized to minimize the rate-distortion cost. A higher skip ratio does not necessarily guarantee higher coding gain.
For example, as shown in Fig. \ref{fig:fre_low} and \ref{fig:fre_high}, at lower bitrates, there are more latents that have negligible bitrates, compared to at higher bitrates. Skipping these latent variables has minimal influence on the overall compression performance. Therefore, even though the skip ratio tends to be higher at lower bitrates than at higher bitrates, the performance improvement achieved by the skip mode remains relatively consistent across different bitrate levels.

\begin{figure}[!t]
  \centering
  
    \subfloat[Encoding time.]
    {  
        \includegraphics[width=0.48\linewidth]{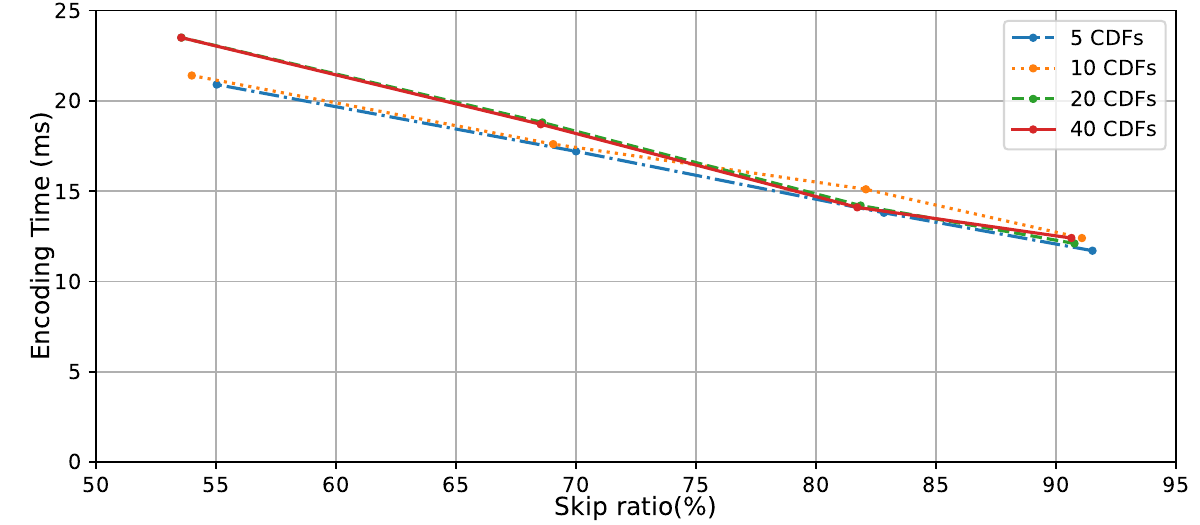}
        \label{fig:enc}
    } 
    \subfloat[Decoding time.]
    {  
        \includegraphics[width=0.48\linewidth]{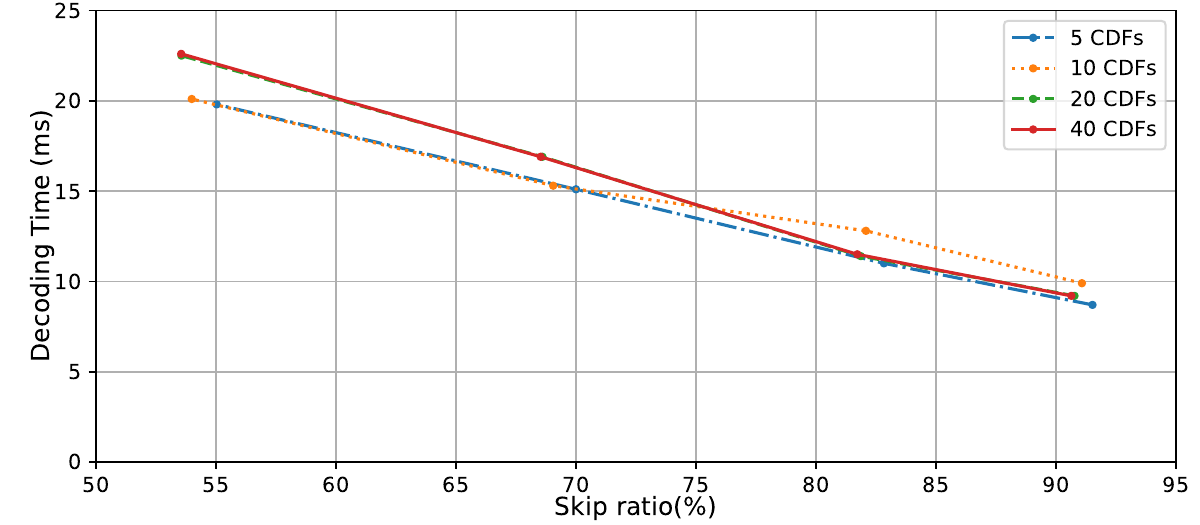}
        \label{fig:dec}
    } 
    
    \caption{
    Relationship between coding time and skip ratio. The results are collected from the Kodak set on the FastNIC model.
    }
    \label{fig:time_skip_ratio}
\end{figure}

Additionally, as reported in Table \ref{tab:enc_time} and \ref{tab:dec_time}, the skip mode significantly reduces coding time. Figure \ref{fig:fre_low} and \ref{fig:fre_high} show that the skip mode eliminates many latent variables with smaller rates. The reduction in the number of coded latents results in shorter coding time, as shown in Table \ref{tab:fastnic_num_cdf} and \ref{tab:fm_num_cdf}. At lower bitrates, the proportion of latent variables with smaller rates increases, leading to a higher skip ratio, as shown in Fig. \ref{fig:skip_bpp}. As shown in Fig. \ref{fig:time_skip_ratio}, both encoding and decoding times decrease as the skip ratio increases. 

\subsection{Analyses on Learned Prior Set}
\begin{figure}[!t]
\centering
\subfloat[Switch w/o skip ($\lambda=0.0018$)]
{  
    \includegraphics[width=0.48\linewidth]{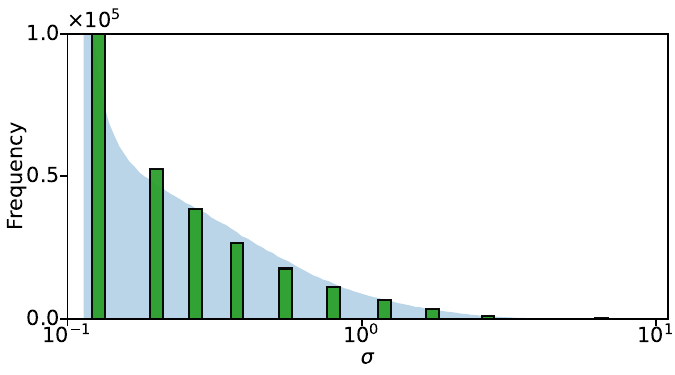}
    \label{fig:gm_wo_skip_0018}
}\hspace{-0.3cm}
\subfloat[Switch w/o skip ($\lambda=0.0483$)]
{  
    \includegraphics[width=0.48\linewidth]{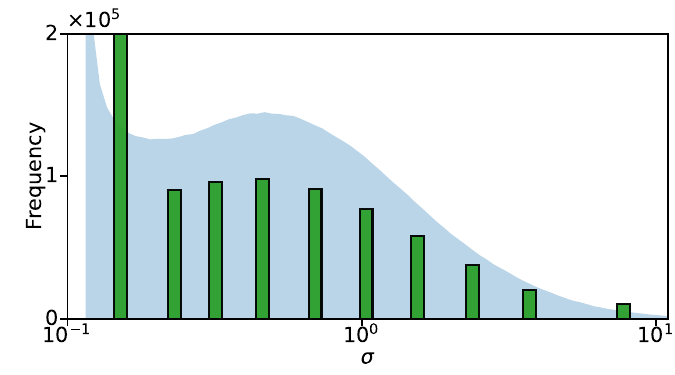}
    \label{fig:gm_wo_skip_0483}
} 

\subfloat[Switch ($\lambda=0.0018$)]
{  
    \includegraphics[width=0.48\linewidth]{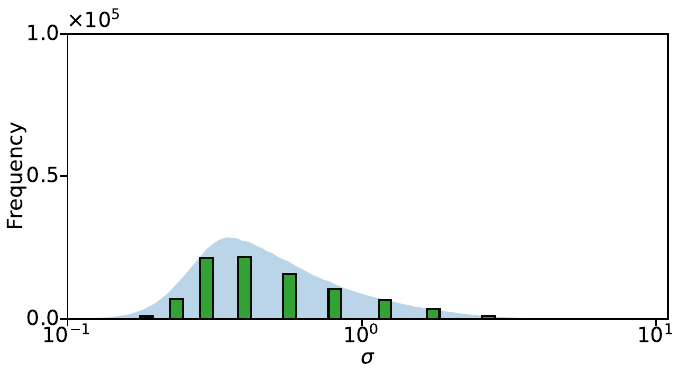}
    \label{fig:gm_w_skip_0018}
}\hspace{-0.3cm}
\subfloat[Switch ($\lambda=0.0483$)]
{  
    \includegraphics[width=0.48\linewidth]{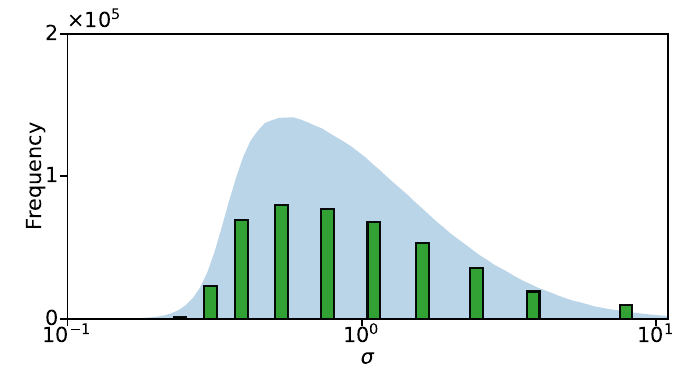}
    \label{fig:gm_w_skip_0483}
} 
%\vspace{-0.1em}
\caption{
Visualization of the learned prior set with \textbf{GM} probabilistic model on the FastNIC model. Each plot shows the frequency of predicted scale values in the original model and the frequency of predicted indexes of priors in the switchable priors method. The results are collected from the Kodak set.
}
\label{fig:gm_frequency}
%\vspace{-0.5em}
\end{figure}

\begin{figure}[!t]
\centering
\subfloat[Switch w/o skip ($\lambda=0.0018$)]
{  
    \includegraphics[width=0.48\linewidth]{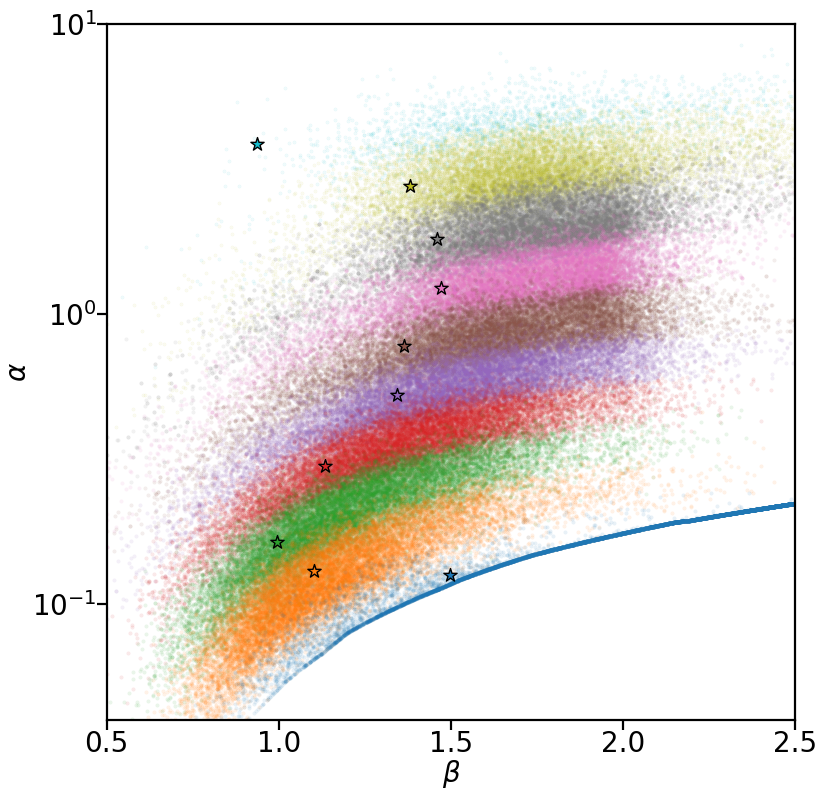}
    \label{fig:ggm_wo_skip_0018}
}
\hspace{-0.3cm}
\subfloat[Switch w/o skip ($\lambda=0.0483$)]
{  
    \includegraphics[width=0.48\linewidth]{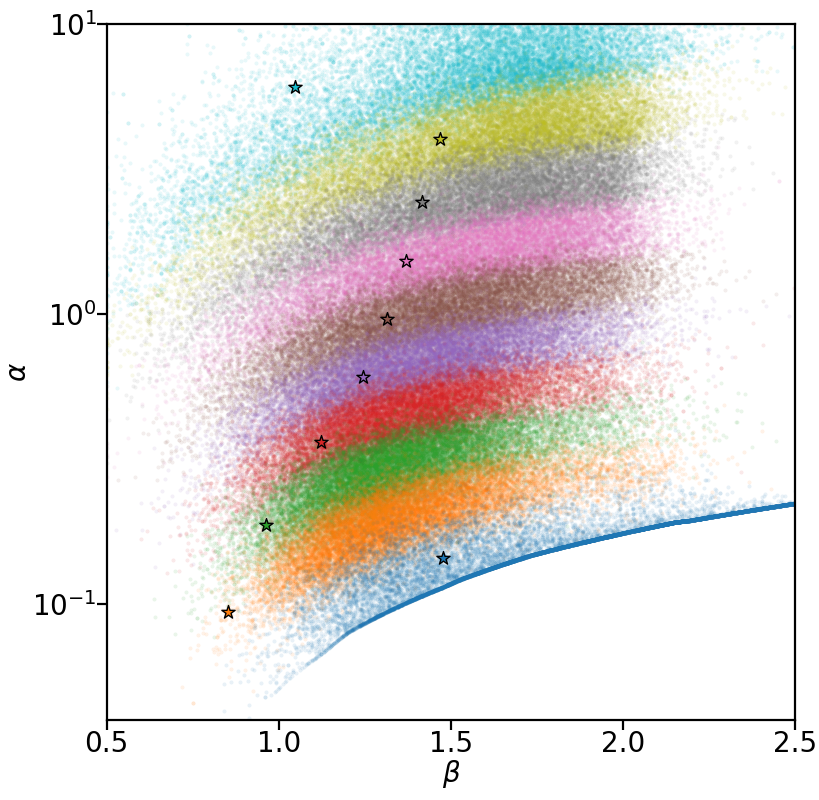}
    \label{fig:ggm_wo_skip_0483}
} 
%\vspace{-0.6em}

\subfloat[Switch ($\lambda=0.0018$)]
{  
    \includegraphics[width=0.48\linewidth]{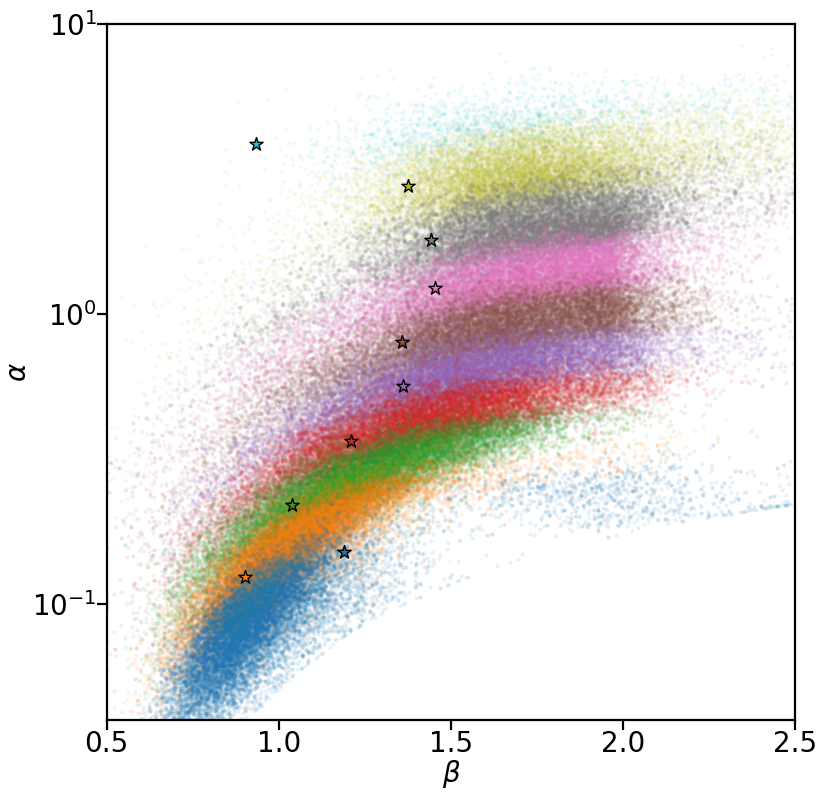}
    \label{fig:ggm_w_skip_0018}
}\hspace{-0.3cm}
\subfloat[Switch ($\lambda=0.0483$)]
{  
    \includegraphics[width=0.48\linewidth]{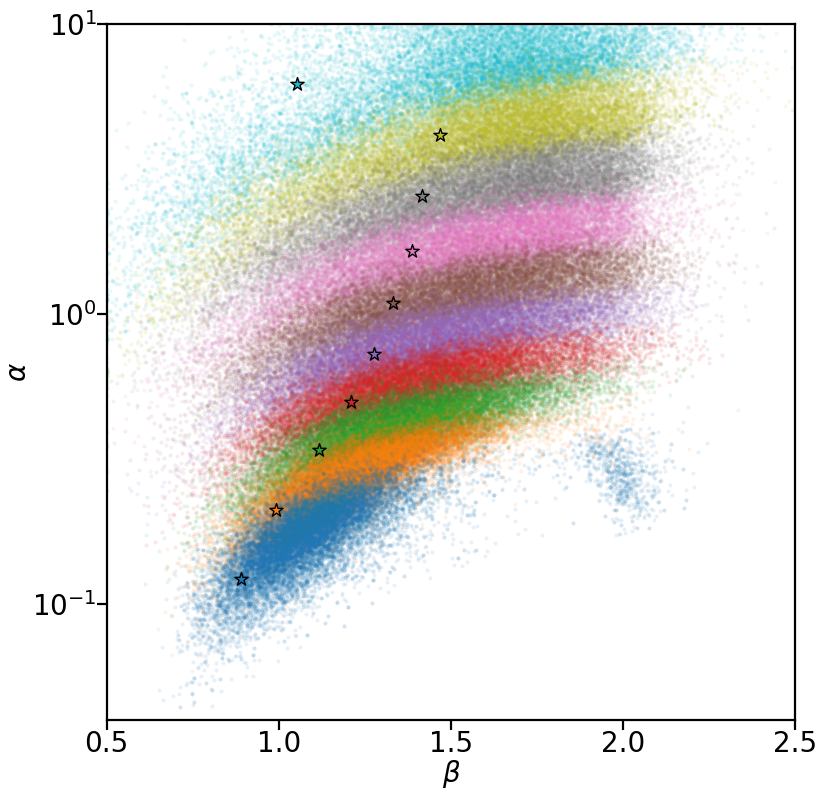}
    \label{fig:ggm_w_skip_0483}
} 
%\vspace{-0.1em}
\caption{
Visualization of the learned prior set with \textbf{GGM} probabilistic model on the FastNIC model. Each plot shows the frequency of predicted $(\beta,\alpha)$ pairs in the original model and the locations of learned distributions in the prior set in our switchable priors method. The locations of learned distributions are marked by $\star$. The entropy parameters predicted from the original entropy model are depicted with different colors and latent variables sharing the same predicted index are marked with the same color. The results are collected from the Kodak set.
}
\label{fig:ggm_scatter}
%\vspace{-1em}
\end{figure}

In this section, we analyze the learned parameterized distributions in the prior set. For simplicity in visualization, we use a learned prior set with a length of 10.

For using GM as the probabilistic model in the prior set, each distribution is parameterized by a scale value $\sigma$. As shown in Fig. \ref{fig:gm_wo_skip_0018} and \ref{fig:gm_wo_skip_0483}, the learned scale values are non-uniformly distributed.
For Switch w/o skip, the frequency of the predicted indexes corresponding to the prior set closely matches the frequency of predicted scale values in the original FastNIC model. After incorporating the skip mode, many latent variables with smaller scale values are skipped, causing the learned scales to shift toward larger values, as shown in Fig. \ref{fig:gm_w_skip_0018} and \ref{fig:gm_w_skip_0483}. This adjustment allows for the use of fewer CDF tables while maintaining performance.

For using GGM as the probabilistic model in the prior set, each distribution is parameterized by a pair of $(\beta, \alpha)$. Figure \ref{fig:ggm_scatter} illustrates the locations of the learned $(\beta, \alpha)$ pairs. In the prior set, the learned shape parameters $\beta$ of the distributions are distinct, highlighting the effectiveness of GGM and the ability of our switchable priors method to support it. Our approach can be viewed as a learning process for soft clustering. In Fig. \ref{fig:ggm_scatter}, entropy parameters predicted by the original entropy model are displayed in different colors, with latent variables sharing the same predicted index marked in the same color. Some mixture points appear at the boundaries between two intervals. This soft assignment arises from the joint training of the prior set and entropy model, indicating that our method adapts to the characteristics of the latents, rather than simply discretizing the original entropy parameters. After incorporating the skip mode, many latents with lower entropy are skipped, causing the learned distributions in the prior set to become more concentrated in regions of higher entropy.

\subsection{Analyses on Different Probabilistic Models}
\label{sec:differmodel}
\begin{table}[!t]
\centering
\caption{Performance of different probabilistic models on \textbf{FastNIC} model.}
\resizebox{\linewidth}{!}{
\begin{threeparttable}
\begin{tabular}{ c|c|ccc|cc} 
\hline
\multirow{2}{*}{Method} &\multirow{2}{*}{\Centerstack[c]{Number\\of CDF}}& \multicolumn{3}{c|}{BD-rate (\%)\tnote{2}} & \multirow{2}{*}{\Centerstack[c]{GGM v.s.\\GM (\%)\tnote{3}}} & \multirow{2}{*}{\Centerstack[c]{GGM v.s.\\GMM (\%)\tnote{3}}} \\
&&GM&GMM&GGM&\\
\hline
Estimated\tnote{1}&-&4.45&2.72 &2.28&-2.08&-0.43 \\
\hline
\multirow{4}{*}{\Centerstack[c]{Switch\\w/o skip}}&5&10.81&7.66 &6.86&-3.56&-0.74 \\
&10&5.78&3.90 &3.44&-2.22&-0.45 \\
&20&4.68&3.00 &2.53&-2.05&-0.46 \\
&40&4.40&2.68 &2.26&-2.05&-0.41 \\
\hline
\multirow{4}{*}{\Centerstack[c]{Switch}}&5&10.28&6.99 &6.17&-3.73&-0.77 \\
&10&5.86&3.97 &3.49&-2.24&-0.46 \\
&20&4.76&2.88 &2.40&-2.25&-0.46 \\
&40&4.14&2.43 &2.00&-2.06&-0.42 \\
\hline
\end{tabular}
\begin{tablenotes}
\footnotesize
\item[1] The estimated rate is calculated through cross-entropy.
\item[2] BD-rate$\downarrow$ is calculated relative to BPG on the USTC-TD \cite{li2024ustc} dataset. All models are trained for MSE and evaluated with PSNR.
\item[3] ``GGM v.s. GM (\%)'' shows the BD-rate gain of GGM compared to GM. ``GGM v.s. GMM (\%)'' shows the BD-rate gain of GGM compared to GMM.
\end{tablenotes} 
\end{threeparttable}
}
\label{tab:compare_gm_gmm_ggm}
\end{table}
In this section, we conducted the performance comparison between GM, GMM, and GGM within the switchable priors framework. First, we analyze the gap between the performance of using switchable priors and the ideal performance, which is evaluated by the estimated rate of the model directly predicting the parameters of the probabilistic models. As shown in Table \ref{tab:compare_gm_gmm_ggm}, Switch w/o skip with only 40 CDF tables achieves performance comparable to the ideal performance for both GM, GMM, and GGM. 

Next, we examine the performance gap between GM, GMM, and GGM under the same network complexity and number of CDF tables. As shown in Table \ref{tab:compare_gm_gmm_ggm}, GGM consistently outperforms GM and GMM under the same complexity constraints. The performance gain from GGM is more significant when the prior set is small, due to its enhanced distribution modeling capability. As the size of the prior set increases, the performance gain of GGM gradually converges to the ideal performance gap.

\subsection{Ablation Studies}
\begin{table}[!t]
\centering
% %\vspace{-0.8em}
\caption{Performance of Top-K acceleration on FastNIC model}
%\vspace{-0.6em}
\begin{threeparttable}
\begin{tabular}{ cc|cc|cc} 
\hline
\multirow{3}{*}{\Centerstack[c]{Method}} &\multirow{3}{*}{\Centerstack[c]{Number\\of CDF}} & \multicolumn{2}{c|}{\Centerstack[c]{Time (s)\tnote{1}}} & \multicolumn{2}{c}{\Centerstack[c]{BD-rate (\%)\tnote{2}}} \\
&&\Centerstack[c]{w/o\\Top-K}&\Centerstack[c]{w/\\Top-K}&\Centerstack[c]{w/o\\Top-K}&\Centerstack[c]{w/\\Top-K}\\
\hline
\multirow{4}{*}{\Centerstack[c]{Switch\\w/o skip}}&5&0.10&0.10&0.03&0.09\\
&10&0.11&0.10&-2.58&-2.66\\
&20&0.13&0.10&-3.32&-3.34\\
&40&0.13&0.10&-3.52&-3.51\\
\hline
\end{tabular}
\begin{tablenotes}
\footnotesize
\item[1] Time in this table is the cost of one iteration during training.
\item[2] BD-rate$\downarrow$ is calculated relative to BPG on Kodak. All models are trained for MSE and evaluated with PSNR.
\end{tablenotes} 
\end{threeparttable}
\label{tab:ablation_topk}
%\vspace{-0.5em}
\end{table}
As shown in Table \ref{tab:ablation_topk}, with the Top-K acceleration, the training process is faster with comparable performance. 
\begin{table}[!t]
\centering
% %\vspace{-0.7em}
\caption{Performance of reusing the prior set on FastNIC model}
%\vspace{-0.6em}
\begin{threeparttable}
\begin{tabular}{ cc|cc} 
\hline
\multirow{2}{*}{\Centerstack[c]{Method}}&\multirow{2}{*}{\Centerstack[c]{Number\\of CDF}}&\multicolumn{2}{c}{\Centerstack[c]{BD-rate (\%)\tnote{1}}}\\
 &&w/o reusing&w/ reusing\\
\hline
\multirow{4}{*}{\Centerstack[c]{Switch\\w/o skip}}&5&-0.02&0.09\\
&10&-2.67&-2.66\\
&20&-3.32&-3.34\\
&40&-3.49&-3.51\\
\hline
\end{tabular}
\begin{tablenotes}
\footnotesize
\item[1] BD-rate$\downarrow$ is calculated relative to BPG on Kodak. All models are trained for MSE and evaluated with PSNR.
\end{tablenotes} 
\end{threeparttable}
\label{tab:ablation_shared}
%\vspace{-0.5em}
\end{table}
Through reusing the prior set for hyperlatents, the CDF tables for coding $\hat{z}$ can be mitigated without obvious influence on the compression performance, as shown in Table \ref{tab:ablation_shared}.
\subsection{Performance of Multi-Dimensional Prior Set for Nonzero-Center Quantization}

Most recent advanced neural image compression models \cite{he2022elic, liu2023learned, li2024fm} adopt zero-center quantization ($\lfloor y-\mu \rceil$) to reduce the influence of train-test mismatch \cite{zhang2023uniform} for better performance. Meanwhile, some studies \cite{cheng2020learned, he2021checkerboard, mentzer2023m2t} adopt nonzero-center quantization ($\lfloor y\rceil$) for parallel training. In the previous sections, we have already demonstrated the effectiveness of the one-dimensional prior set for zero-center quantization. In this section, we show the necessity of the multi-dimensional prior set for nonzero-center quantization.

\begin{figure}[!t]
\centering
\subfloat[1-d prior set]
{  
    \includegraphics[width=0.47\linewidth]{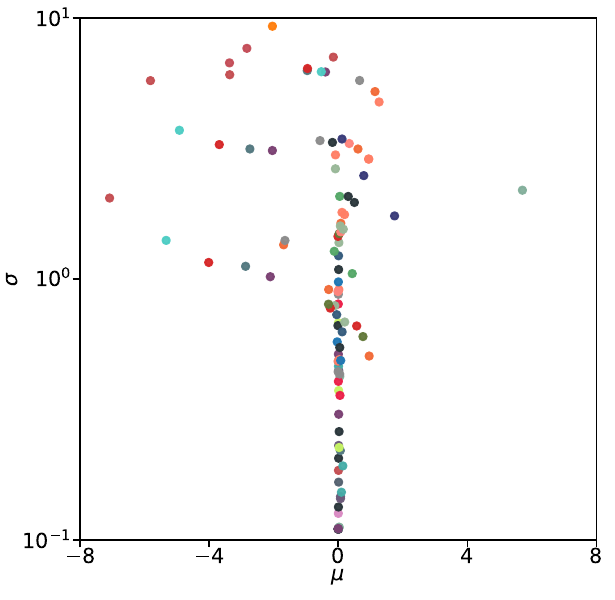}
    \label{fig:1d_gmm}
}\hspace{-0.3cm}
\subfloat[2-d prior set]
{  
    \includegraphics[width=0.47\linewidth]{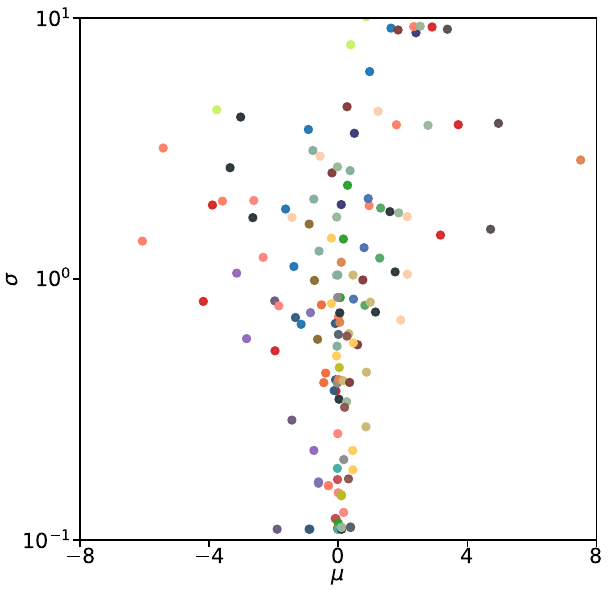}
    \label{fig:2d_gmm}
} 
%\vspace{-0.1em}
\caption{ Visualization of learned prior set of GMM on Cheng-ckbd. Each plot shows the location of different $(\mu,\sigma)$ pairs of GMM in the learned prior set. There are 40 distributions in the 1-d prior set, containing 120 $(\mu,\sigma)$ pairs. There are 50 distributions in the 2-d prior set, containing 150 $(\mu,\sigma)$ pairs. Pairs from the same GMM distribution are marked with the same color.
}
\label{fig:gmm}
%\vspace{-0.8em}
\end{figure}

\begin{table}[!t]
\centering
\caption{Compression performance of different methods on \textbf{Cheng-ckbd} model.}
%\vspace{-0.3em}
\begin{threeparttable}
\begin{tabular}{ c|ccc|c} 
\hline
\multirow{2}{*}{Method} & \multicolumn{3}{c|}{Number of CDF\tnote{1}} & \multirow{2}{*}{BD-rate (\%)\tnote{2}} \\
&dim1&dim2&All&\\
\hline
\multirow{1}{*}{\Centerstack[c]{Dynamically computed CDF}}&-&-&294912&0.00\\
\hline
\multirow{4}{*}{\Centerstack[c]{Switch 1d\\w/o skip}}&-&-&5&10.75\\
&-&-&10&7.08\\
&-&-&20&5.81\\
&-&-&40&5.58\\
\hline
\multirow{8}{*}{\Centerstack[c]{Switch 2d\\w/o skip}}&5&5&25&3.58\\
&10&5&50&2.51\\
&5&10&50&1.44\\
&10&10&100&0.83\\
&5&20&100&0.81\\
&10&20&200&0.04\\
&5&40&200&0.45\\
&10&40&400&-0.25\\
\hline
\multirow{8}{*}{\Centerstack[c]{Switch 2d}}&5&5&25&3.33\\
&10&5&50&2.47\\
&5&10&50&1.38\\
&10&10&100&0.80\\
&5&20&100&0.70\\
&10&20&200&-0.10\\
&5&40&200&0.38\\
&10&40&400&\textbf{-0.36}\\
\hline
\end{tabular}
\begin{tablenotes}
\footnotesize
\item[1] For 2d prior set, we initialize dim1 of the prior set with different $\mu$ and initialize dim2 with different $\sigma$ for GMM to stabilize training. 
\item[2] BD-rate is compared to the performance of dynamically computing CDF on the Kodak dataset. All models are trained for MSE and evaluated with PSNR.
\end{tablenotes} 
\end{threeparttable}
\label{tab:gmm}
%\vspace{-0.8em}
\end{table}

As shown in Fig. \ref{fig:1d_gmm}, for the Cheng-ckbd model, which employs nonzero-center quantization and GMM, the one-dimensional prior set reveals its limitations. Specifically, the distributions learned in the prior set have similar mean values when scale parameters are small, leading to a notable performance drop, as shown in Table \ref{tab:gmm}. This limitation suggests that the one-dimensional prior set struggles to jointly capture both the mean and scale characteristics.
In contrast, the two-dimensional prior set effectively addresses this challenge. As illustrated in Fig. \ref{fig:2d_gmm}, it successfully captures both the mean and scale characteristics of the distributions. Moreover, as shown in Table \ref{tab:gmm}, the two-dimensional prior set significantly outperforms the one-dimensional prior set, even when both use the same number of CDF tables. Our experimental results also demonstrate that even when the probabilistic model is GMM with a nonzero mean, the two-dimensional prior set is sufficiently capable of handling such cases effectively.

\section{Conclusion}
In this paper, we propose a method to learn switchable priors, effectively decoupling the complexity of neural image compression models from the adopted probabilistic model. Our approach constructs a trainable prior set containing various parameterized distributions. Instead of directly estimating the parameters of the probabilistic model, the entropy model predicts the index of the most suitable distribution in the prior set. This allows for the adoption of more complex probabilistic models without increasing complexity. To further enhance compression performance and reduce complexity, we incorporate a skip mode into the prior set. Moreover, we introduce a lightweight neural image compression model combined with the proposed switchable priors method, which outperforms BPG in compression performance with encoding and decoding complexities below 12 KMACs/pixel and 10 KMACs/pixel, respectively. We conduct extensive experiments to demonstrate the effectiveness of the proposed switchable priors method, showing significant reductions in entropy coding complexity and notable improvements in compression performance.
\bibliographystyle{IEEEtran}
\bibliography{main}
\newpage

\appendices

\section{Comparison of skipping mechanism}
First, we summarize the implementation of previous approaches. The skipping mechanism proposed in \cite{shi2022alphavc} is designed for the Gaussian model. In this approach, during inference, latent variables whose scale parameters fall below a predefined threshold are skipped. Since this method relies on the Gaussian scale parameter, it cannot be directly applied to the generalized Gaussian model or other probabilistic models. 

In \cite{lee2022selective}, the authors introduced additional networks to predict a binary mask indicating which latents should be skipped. These networks are jointly trained with the entire neural image compression model. Due to the presence of quantization approximations (such as adding uniform noise) during joint training, the skipping mechanism is optimized to minimize a proxy objective rather than the true rate-distortion objective using rounding. During the training process in \cite{lee2022selective}, the forward process of the mask generation during training can be formulated as
\begin{equation}
\tilde{b} = B(b + U(-0.5, 0.5)),
\end{equation}
where $b$ is the importance map predicted by the entropy model and clamped into $(0,1)$, $B(\cdot)$ is a rounding operator, $\tilde{b}$ is the stochastic binary mask used during training.
To enable gradient propagation through the binary mask generation process, they employed a simple straight-through estimator, allowing gradients to flow through the rounding operator as if the rounding operation were an identity function.

Then, we discuss how our approach differs from previous approaches. Similar to \cite{lee2022selective}, our skipping mechanism also utilizes additional networks to predict the binary mask. However, we have developed more effective training methods to enhance performance. 
\begin{itemize}
    \item First, we optimize the additional network using the true rate-distortion objective estimated by rounded latents rather than the proxy objective caused by quantization approximation during training in \cite{lee2022selective}. Specifically, we separately train the additional network based on a well-trained neural image compression model. During the optimization of the additional network, all other modules in the neural image compression model are kept fixed. This design allows us to use rounded latents to estimate the rate-distortion cost, thereby optimizing the true objective.
    \item Second, we employ the Gumbel-softmax \cite{jang2017gumbel} estimator to enable gradient propagation through the binary mask generation process. This method provides a more accurate gradient estimation compared to the simple straight-through estimator in \cite{lee2022selective}.
\end{itemize}
These enhancements result in improved performance of our skipping mechanism over previous approaches. Additionally, compared to the skipping approaches in \cite{shi2022alphavc}, our method is not limited to the Gaussian model, making it more general and applicable to a wider range of scenarios.

We demonstrate the effectiveness of our skipping mechanism through experimental analyses, as shown in Tables \ref{tab:fast_skip} and \ref{tab:fm_skip}. Since the method described in \cite{shi2022alphavc} is designed for the Gaussian model, we evaluate the performance of our skipping mechanism using the same setting, \textit{i.e.}, use the Gaussian model as the probabilistic model. As shown in Tables \ref{tab:fast_skip} and \ref{tab:fm_skip}, with an appropriate skip threshold, the method in \cite{shi2022alphavc} can reduce coding time without obviously influencing compression performance. For the method described in \cite{lee2022selective}, it improves compression performance on FastNIC but shows worse performance on the more advanced FM-intra. This discrepancy may be due to gradient estimation errors introduced by the straight-through estimator used in their approach.

With our advanced training strategy, which includes optimizing the additional network using the true rate-distortion objective and employing the Gumbel-softmax estimator, our method achieves both improved compression performance and comparable coding time compared to \cite{lee2022selective}. When compared to \cite{shi2022alphavc}, our method also demonstrates better compression performance and shorter coding time. 

These results highlight the advantages of our proposed skipping mechanism, particularly its ability to enhance both coding efficiency and compression performance.

\section{LUTs-based implementation for entropy coding} 

First, we introduce the overview of the look-up tables-based (LUTs-based) method for entropy coding. The LUTs-based method only influences the test stage and does not influence the training process. After training, several entropy parameter values of the probabilistic model are sampled in a specific manner from their respective possible ranges. We then calculate the corresponding cumulative distribution function (CDF) table for each sampled value. The encoder and decoder share these pre-computed CDF tables. During the actual encoding and decoding process, the predicted entropy parameters, which model the distribution of latent variables, are quantized to the nearest sampled value. These quantized values are then used to index the corresponding CDF table, which is necessary for entropy coders. For both the Gaussian model (GM) and the generalized Gaussian model (GGM), which have an identifiable mean parameter, we follow previous studies \cite{minnen2020channel,he2022elic,liu2023learned} to apply zero-center quantization, where the symbols encoded into bitstreams are given by $\lfloor y - \mu \rceil$. For GM, the entropy parameter involved in entropy coding is $\sigma$, while for GGM, the entropy parameters involved in entropy coding are $\beta$ and $\alpha$.

Then, we introduce the sampling strategy and entropy coding process for GM and GGM.
For GM, we follow previous studies \cite{begaint2020compressai} to linearly sample $M$ values in the log-scale field of $[0.11,60]$ for the $\sigma$ parameter. The sampling strategy is
\begin{align}
    \sigma_i = \text{exp}\left(\text{log}(0.11)+i\times\frac{\text{log}(60)-\text{log}(0.11)}{M-1}\right),
\end{align}
where $i\in\{0,1,\cdots,M-1\}$. We then calculate the corresponding CDF table for each $\sigma_i$. The precision of the CDF table is 16-bit unsigned integer (uint16) with a maximum length of 256 per table. The performance with different numbers of CDF tables is included in the manuscript. These CDF tables are stored on both the encoder and decoder sides for entropy coding. During the test stage, the $\sigma$ parameter predicted by the entropy models is quantized to the nearest sampled value, which is then used to index the corresponding CDF table for entropy coding.

\begin{table}[!t]
\centering
\caption{Performance comparison of different skipping mechanism on our \textbf{FastNIC} with Gaussian model.}
\begin{threeparttable}
\begin{tabular}{ c|cc|cc|c}
\hline
\multicolumn{3}{c|}{\multirow{2}{*}{\Centerstack[c]{Method}}} & \multicolumn{2}{c|}{\Centerstack[c]{Time (ms)}}  & \multirow{2}{*}{\Centerstack[c]{BD-rate (\%)\tnote{2}}}\\
\multicolumn{3}{c|}{}&Enc.\tnote{1}&Dec.\tnote{1}&\\
\hline
\multicolumn{3}{c|}{\Centerstack[c]{w/o skip mode}}&37.7 &35.7 &0.00 \\
\hline
\multirow{9}{*}{\Centerstack[c]{w/ skip mode}}&\multirow{7}{*}{\Centerstack[c]{Shi2022\tnote{3}\\\cite{shi2022alphavc}}}&0.06 &32.8 &32.0 &-0.01 \\
&&0.08 &32.1 &31.2 &0.01 \\
&&0.10 &30.2 &29.1 &0.00 \\
&&0.12 &22.3 &21.2 &-0.07 \\
&&0.14 &21.5 &20.2 &0.05 \\
&&0.17 &20.5 &19.0 &0.15 \\
&&0.20 &19.9 &18.4 &0.18 \\
\cline{2-6}
&\multicolumn{2}{c|}{\Centerstack[c]{Lee2022 \cite{lee2022selective}}}&15.9 &13.2 &-0.51 \\
\cline{2-6}
&\multicolumn{2}{c|}{\Centerstack[c]{Ours}}&16.1 &13.6 &\textbf{-0.86} \\
\hline
\end{tabular}
\begin{tablenotes}
\footnotesize
\item[1] Coding times and compression performance are evaluated on the Kodak dataset. ``Encoding" is abbreviated as ``Enc." ``Decoding" is abbreviated as ``Dec."
\item[2] BD-rate$\downarrow$ is compared to the model without skip mode. All models are trained for MSE and evaluated with PSNR.
\item[3] The Shi2022 method proposed in \cite{shi2022alphavc} introduces a threshold to determine which latent variables should be skipped, based on the scale parameters estimated by the entropy model. The results with different thresholds are shown in this table.
\end{tablenotes} 
\end{threeparttable}
\label{tab:fast_skip}
\end{table}

\begin{table}[!t]
\centering
\caption{Performance comparison of different skipping mechanism on \textbf{FM-intra} \cite{li2024fm}  with Gaussian model}
\begin{threeparttable}
\begin{tabular}{ c|cc|cc|c}
\hline
\multicolumn{3}{c|}{\multirow{2}{*}{\Centerstack[c]{Method}}} & \multicolumn{2}{c|}{\Centerstack[c]{Time (ms)}}  & \multirow{2}{*}{\Centerstack[c]{BD-rate (\%)\tnote{2}}}\\
\multicolumn{3}{c|}{}&Enc.\tnote{1}&Dec.\tnote{1}&\\
\hline
\multicolumn{3}{c|}{\Centerstack[c]{w/o skip mode}}&64.6 &67.6 &0.00 \\
\hline
\multirow{9}{*}{\Centerstack[c]{w/ skip mode}}&\multirow{7}{*}{\Centerstack[c]{Shi2022\tnote{3}\\\cite{shi2022alphavc}}}&0.06 &66.2 &71.6 &0.01 \\
&&0.08 &65.7 &71.2 &0.01 \\
&&0.10 &60.8 &66.2 &-0.01 \\
&&0.12 &45.0 &48.1 &-0.03 \\
&&0.14 &44.2 &47.0 &-0.05 \\
&&0.17 &43.7 &46.3 &-0.04 \\
&&0.20 &43.0 &45.2 &0.01 \\
\cline{2-6}
&\multicolumn{2}{c|}{\Centerstack[c]{Lee2022 \cite{lee2022selective}}}&29.5 &30.3 &5.23 \\
\cline{2-6}
&\multicolumn{2}{c|}{\Centerstack[c]{Ours}}&31.8 &32.1 &\textbf{-0.15} \\
\hline
\end{tabular}
\begin{tablenotes}
\footnotesize
\item[1] Coding times and compression performance are evaluated on the Kodak dataset. ``Encoding" is abbreviated as ``Enc." ``Decoding" is abbreviated as ``Dec."
\item[2] BD-rate$\downarrow$ is compared to the model without skip mode. All models are trained for MSE and evaluated with PSNR.
\item[3] The Shi2022 method proposed in \cite{shi2022alphavc} introduces a threshold to determine which latent variables should be skipped, based on the scale parameters estimated by the entropy model. The results with different thresholds are shown in this table.
\end{tablenotes} 
\end{threeparttable}
\label{tab:fm_skip}
\end{table}

\begin{table}[!t]
\centering
\caption{Performance of LUTs-based method for GGM on \textbf{FastNIC} model}
\begin{threeparttable}
\begin{tabular}{cccccc}
\hline
\Centerstack[c]{$\beta$\tnote{1}} & \Centerstack[c]{$\alpha$\tnote{1}}& \Centerstack[c]{Number of CDF\tnote{2}} & \Centerstack[c]{BD-rate (\%)\tnote{3}} &\Centerstack[c]{$t_e$ (ms)\tnote{4}} & \Centerstack[c]{$t_d$ (ms)\tnote{4}}\\
\hline
5&5&25&124.73&32.4&34.7\\
\rowcolor{gray!20}
5&10&50&19.30&33.0&34.9\\
5&20&100&2.10&32.9&34.5\\
\rowcolor{gray!20}
5&40&200&-0.97&35.0&35.2\\
5&80&400&-1.69&34.1&36.1\\
5&160&800&-1.86&37.2&38.5\\
10&5&50&122.47&33.7&36.7\\
10&10&100&17.55&33.7&36.6\\
10&20&200&0.50&34.0&35.9\\
10&40&400&-2.37&34.8&36.6\\
\rowcolor{gray!20}
10&80&800&-3.07&35.0&37.4\\
10&160&1600&-3.21&36.5&38.0\\
20&5&100&122.08&33.7&37.1\\
20&10&200&17.21&33.5&36.1\\
20&20&400&0.21&33.1&34.8\\
20&40&800&-2.63&33.3&35.3\\
20&80&1600&-3.32&36.0&37.7\\
\rowcolor{gray!20}
20&160&3200&-3.46&37.4&38.6\\
40&5&200&121.99&33.8&37.4\\
40&10&400&17.14&32.4&35.1\\
40&20&800&0.15&34.0&36.4\\
40&40&1600&-2.69&34.0&36.0\\
40&80&3200&-3.37&35.8&37.2\\
40&160&6400&-3.49&38.6&39.7\\
80&5&400&121.97&33.4&36.9\\
80&10&800&17.12&33.3&35.9\\
80&20&1600&0.14&33.7&35.7\\
80&40&3200&-2.70&35.5&37.1\\
80&80&6400&-3.38&37.4&38.9\\
\rowcolor{gray!20}
80&160&12800&-3.52&43.5&44.6\\
160&5&800&121.96&34.2&37.6\\
160&10&1600&17.12&33.7&36.8\\
160&20&3200&0.14&34.9&36.8\\
160&40&6400&-2.70&38.3&40.0\\
160&80&12800&-3.38&42.3&43.3\\
160&160&25600&-3.53&62.5&57.9\\
\hline
\end{tabular}
\begin{tablenotes}
\footnotesize
\item[1] The results shown in the manuscript are marked with color.
\item[2] The $\beta$ and $\alpha$ columns list the number of samples for shape and scale parameters when constructing CDF tables, respectively.
\item[3] BD-rate$\downarrow$ is evaluated relative to BPG on the Kodak set.
\item[4] $t_e$ and $t_d$ represent the encoding and decoding time, respectively. The time presented in this table includes entropy coding time.
\end{tablenotes} 
\end{threeparttable}
\label{tab:lut_ggm}
\end{table}

For GGM, we follow the implementation in \cite{zhang2024ggm}. The $\beta$ parameter is linearly sampled from the range $[0.5, 3]$ with $N$ samples, and the $\alpha$ parameter is linearly sampled in the log-scale range $[0.01, 60]$ with $M$ samples. Specifically, the sample values of the $\alpha$ parameter are 
\begin{align}
\label{eq:ggm_sample}
    \alpha_i = \text{exp}\left(\text{log}(0.01)+i\times\frac{\text{log}(60)-\text{log}(0.01)}{M-1}\right),
\end{align}
where $i\in\{0,1,\cdots,M-1\}$. We then combine the $N$ $\beta$ values and $M$ $\alpha$ values to generate $M \times N$ $\beta-\alpha$ pairs and calculate the corresponding CDF table for each pair. The precision of the CDF table is set to 16-bit unsigned integer (uint16), with a maximum length of each CDF table set to 256. These pre-computed CDF tables are stored on both the encoder and decoder sides for entropy coding. During the encoding and decoding process, the predicted $\beta$ and $\alpha$ parameters from the entropy models are quantized to the corresponding intervals, respectively, which are then used to index the corresponding CDF tables. Table \ref{tab:lut_ggm} presents the performance of the LUTs-based implementation for GGM with varying numbers of samples.

For the Gaussian mixture model (GMM), which has 9 parameters, using the LUTs-based method would incur excessive storage costs. For example, sampling 20 values for each of the 9 parameters in GMM would require storing $20^9$ CDF tables, resulting in a storage requirement of at least $2 \times 10^5$ GB, which is not practical. Therefore, we only compare our method with calculating the CDF tables dynamically during encoding and decoding for GMM \cite{cheng2020learned, fu2023learned}.

\section{Zero-center quantization for GMM}
Previous study \cite{zhang2023uniform} suggested that zero-center quantization can enhance performance when using the mixed quantization surrogate \cite{minnen2020channel} during training and testing. However, zero-center quantization cannot be directly applied to GMM. To address this limitation and improve the performance of GMM, we designed a zero-center quantization method specifically for GMM. To illustrate our approach, we use a GMM with three Gaussian components as an example. 
For a Gaussian mixture model with three Gaussian components, the predicted entropy parameters are $\{p_1, \mu_1, \sigma_1, \dots, p_3, \mu_3, \sigma_3\}$, where $p_i$, $\mu_i$, and $\sigma_i$ denote the weight, mean, and scale parameters of the $i$-th Gaussian component, respectively. We define the ``mean" of GMM as $\mu = \sum_{i=1}^{3}p_i \mu_i$. After obtaining $\mu$, we apply zero-center quantization, formulated as $\hat{y}=\lfloor y-\mu \rceil+\mu$. 

\vfill
\end{document}